\pdfoutput=1
\documentclass[prb,%
reprint,
superscriptaddress,
nofootinbib,
amsmath,amssymb,
aps,
longbibliography
]{revtex4-1}
\usepackage{graphicx}% Include figure files
\usepackage{dcolumn}% Align table columns on decimal point
\usepackage{bm}% bold math
\usepackage{subcaption}
\usepackage{caption}
\usepackage{float}
\usepackage{lipsum}
\usepackage{amsmath}
\usepackage{amssymb}
\usepackage{dsfont}
\usepackage{color}
\usepackage{nccmath}

\usepackage[,%Uncomment any one of the following lines to test 
text={7.3in,10in},%centering,
]{geometry}

\newenvironment{mpmatrix}{\begin{scriptsize}\begin{pmatrix}}%
		{\end{pmatrix}\end{scriptsize}}%

\begin{document}
	
	%\preprint{APS/123-QED}
	
	\title{Fate of a topological invariant for correlated lattice electrons at finite temperature}% Force line breaks with \\
	
	\author{A.A. Markov}
		\affiliation{
		  Russian Quantum Center, 100 Novaya St., Skolkovo, Moscow 143025, Russia
	}
	\affiliation{Department of Physics, Lomonosov Moskow State University, Leninskie gory 1, 119991 Moscow, Russia}
	\author{G. Rohringer}%
		\affiliation{
		 Russian Quantum Center, 100 Novaya St., Skolkovo, Moscow 143025, Russia
	}
	
	\author{A.N. Rubtsov}
	\affiliation{
		 Russian Quantum Center, 100 Novaya St., Skolkovo, Moscow 143025, Russia
	}
	\affiliation{Department of Physics, Lomonosov Moskow State University, Leninskie gory 1, 119991 Moscow, Russia}

	\date{\today}% It is always \today, today,
	
	\begin{abstract}
Electrons on a two-dimensional (2$d$) lattice which is exposed to a strong uniform magnetic field show intriguing physical phenomena. The spectrum of such systems exhibits a complex (multi-)band structure known as Hofstadter's butterfly. For fillings at which the system is a band insulator one observes a quantized integer-valued Hall conductivity $\sigma_{xy}$ corresponding to a topological invariant, the first Chern number $\mathcal{C}_1$. This is robust against many-body interactions as long as no changes in the gap structure occur. Strictly speaking, this stability holds only at zero temperatures $T$ while for $T>0$ correlation effects have to be taken into account. In this work, we address this question by presenting a dynamical mean field theory (DMFT) study of the Hubbard model in a uniform magnetic field. The inclusion of local correlations at finite temperature leads to (i) a shrinking of the integer plateaus of $\sigma_{xy}$ as a function of the chemical potential and (ii) eventually to a deviation from these integer values. We demonstrate that these effects can be related to a correlation-driven narrowing and filling of the band gap, respectively. 
		\end{abstract}
	
	\maketitle
	
	%\tableofcontents
	
	\section{\label{intro}Introduction}
	
	Topological states of matter and strongly correlated electron systems are two of the most intriguing subjects in frontier condensed matter research. The increasing interest in materials which are featuring such properties is driven by their fascinating physical phenomena which are potentially also of high technological relevance. More specifically, physical properties emerging from a topologically non-trivial band structures are typically very stable against perturbations due to topological protection. Correlation effects on the other hand are responsible for some of the most fascinating physical phenomena such as the Mott metal-to-insulator transition \cite{McWhan1970,McWhan1973} or high-temperature superconductivity \cite{bednorz1986possible}. A physical observable which is of equally high importance in both research areas is the current response function perpendicular to an applied electric field $E$ in the presence of a transverse magnetic field $B$, i.e., the Hall conductivity $\sigma_{xy}$.
	
	Experimentally, it has been shown \cite{Klitzing1980} that $\sigma_{xy}$ of a 2$d$ electron gas exhibits plateaus as a function of $B$ which correspond to integer multiples of the flux quantum $\Phi_0$. This can be understood in terms of filled Landau levels for an electron in a uniform magnetic field. In a lattice system, these Landau levels acquire a finite dispersion (i.e., momentum dependence) which gives rice to a complex band structure known as the Hofstadter butterfly \cite{hofstadter1976energy}. Remarkably, in this situation $\sigma_{xy}$ corresponds to a topological invariant of this band structure, namely, the integral over its Berry curvature\cite{Berry1984} equal to the first Chern number \cite{thouless1982quantized,simon1983holonomy} $\mathcal{C}_1$ which makes this observable robust against small perturbations.
	
	In the presence of correlations on the other hand, an unexpected behavior of $\sigma_{xy}$ as a function of temperature and/or density of charge carries has been found. For instance, for electron-doped high-temperature superconducting cuprates a temperature-dependent sign change occurs in the Hall conductivity \cite{hagen1993anomalous,Jenkins2010} for parameter regimes where the spectral function indicates a purely hole-like Fermi surface. These findings have inspired a number of theoretical studies \cite{castillo1992hall,voruganti1992conductivity,assaad1995hall,shastry1993faraday} of the Hall coefficient at small magnetic fields in one of the most basic models for electronic correlations, i.e., the Hubbard Hamiltonian \cite{Gutzwiller1963,Kanamori1963,Hubbard1963}. It has been demonstrated \cite{rojo1993sign} that the anomalous sign change in $\sigma_{xy}$ occurs indeed in this model system close to half-filling and is purely due to correlation effects. The interplay between correlations and the above mentioned topology behind $\sigma_{xy}$ has been, however, not discussed so far.
	
	%In resent times, high-$T_c$ superconductivity\cite{bednorz1986possible} and topological states of matter\cite{bernevig2013topological} have been attracted increasing interest of the scientific community. For both these fields of research the study of the Hall effect in a system with the Hubbard on-site interaction, which is the aim of our paper, is interesting. On the one hand, it would deepen our understanding of the Hubbard model, which is believed to describe essentially important physics of high-Tc cuprates \cite{anderson1997}. On the other hand, two-dimensional lattice electrons subjected to a uniform magnetic field (Hofsdater model \cite{hofstadter1976energy}) is one of the first known examples of topologically non-trivial phase of matter featuring with the quantized Hall conductivity\cite{thouless1982quantized} at zero temperature. Therefore, consideration of the system with the Hubbard interaction and at non-zero temperatures would provide us with new insights about the stability of the Hall conductivity and its limitations.
	                
	This question has, in fact, stimulated several studies for the Hall and the Spin Hall conductivity for other topologically non-trivial systems (for a recent review see \cite{hohenadler2013correlation} or \cite{rachel2018interacting}) especially after Kane-Mele's seminal paper \cite{kane2005quantum} where a topologically nontrivial time-reversal model has been introduced. The effect of Hubbard interactions on two-dimensional topological matter was studied in many possible settings including the Kane-Male-Hubbard model \cite{zheng2011particle,wu2012quantum} or the Bernevig-Hughes-Zhang-Hubbard (BHZH) model \cite{yoshida2012correlation,budich2013fluctuation} (more references in \cite{rachel2018interacting}). Different approaches of treating the interaction term including cluster extensions of the Dynamical Mean Field Theory (DMFT) \cite{wu2012quantum} or numerically exact Quantum Monte Carlo calculations \cite{hohenadler2014phase,budich2013fluctuation} have been used to calculate phase diagrams for these models. The temperature dependence of the Hall conductivity has been addressed in the BHZH model \cite{yoshida2012correlation}. There, it was found that due to interaction effects quasiparticle peaks are getting closer (corresponding to a narrowing of the gap) which renders the Spin-Hall Conductivity more sensitive to changes in the temperature. Another example of a topological system is the time-reversal invariant  Hofsdatdter model which has been studied with negative and positive Hubbard interaction \cite{cocks2012time,wang2014topological,kumar2016interaction}. The probably most natural framework for the analysis of the Hall conductivity $\sigma_{xy}$ is the Hubbard model in a finite magnetic field, i.e., the standard Hofsdater model with broken time-reversal symmetry, which has --to the best of our knowledge-- not been addressed so far.
	
	In this paper, we fill this gap by presenting a study of the Hall effect in the Hubbard-Hofstadter model. Using  the dynamical mean field theory (DMFT) approach we show that a combined effect of the Hubbard interaction and finite temperatures strongly affects the Hall conductivity and eventually leads to a breakdown of the integer quantum Hall regime. Moreover, we are able to recover the above discussed change of sign of the Hall conductivity close to half-filling.
	
	The plan of our paper is the following: In Sec.~\ref{Ham}, we introduce the 2$d$ Hubbard model in a uniform magnetic field. In Sec.~\ref{top}, we discuss the Hall conductivity $\sigma_{xy}$ at zero temperature from a topological perspective, while in Sec.~\ref{DMFT} we address the most general situation of finite temperatures and interactions in the framework of DMFT. In Sec.~\ref{results} we present our numerical results and Sec.~\ref{conclusions} is devoted to conclusions and an outlook. 
	
	%$\bullet$ The Quantum Hall effect.

	%$\bullet$ Correlations and topological states of matter.
	
	%$\bullet$ Hubbard model in a magnetic field.  

	\section{\label{Model}Model and methods}
	
	%$\bullet$ Hofstadter model and magnetic bands. TRB.%
	
	\subsection{\label{Ham}Hubbard Hamiltonian in a magnetic field}
	
 	We consider the Hubbard model \cite{Hubbard1963} on a two-dimensional square lattice (in the $xy$ plane) with a lattice constant $a\!=\!1$ in a uniform magnetic field $\mathbf{B}\!=\!B\mathbf{e_z}$ in $z$ direction (i.e., perpendicular to  the $2d$ lattice):
	
	\begin{equation}
	\label{ham}
	H = - \sum_{\langle i,j\rangle,\sigma}\left(t_{ij} c^{\dagger}_{i\sigma} c_{j\sigma}+h.c.\right) + \sum_{i}U n_{i\uparrow}n_{i\downarrow},
	\end{equation}
	where $c^{(\dagger)}_{i\sigma}$ annihilates (creates) an electron with spin $\sigma$ at the lattice site $\mathbf{R}_i$, $n_{i\sigma}\!=\!c^{\dagger}_{i\sigma}c_{i\sigma}$ and $U\!>\!0$ denotes the constant Coulomb repulsion between two electrons at the same lattice site. The magnetic field is coupled to the system by means of the Peierl's substitution \cite{Peierls1933} which introduces a site-dependent phase factor in the hopping matrix $t_{ij}$
	
	\begin{equation}
	t_{ij} = t\cdot e^{i\frac{2\pi}{\Phi_0}\int_{\bm{R}_i}^{\bm{R}_j}\bm{A}(\bm{r})\cdot \bm{dr}},
	\label{hop}
	\end{equation}

	where $t$ denotes the hopping amplitude between neighboring sites $\langle ij\rangle$. $\mathbf{A}(\mathbf{r})\!=\!-B(y,0,0)$ is the vector potential in the Landau gauge and $\Phi_0\!=\!\frac{h}{e}$ corresponds to the flux quantum.
	
	\begin{figure}
		\centering
		\includegraphics[width=0.5\linewidth]{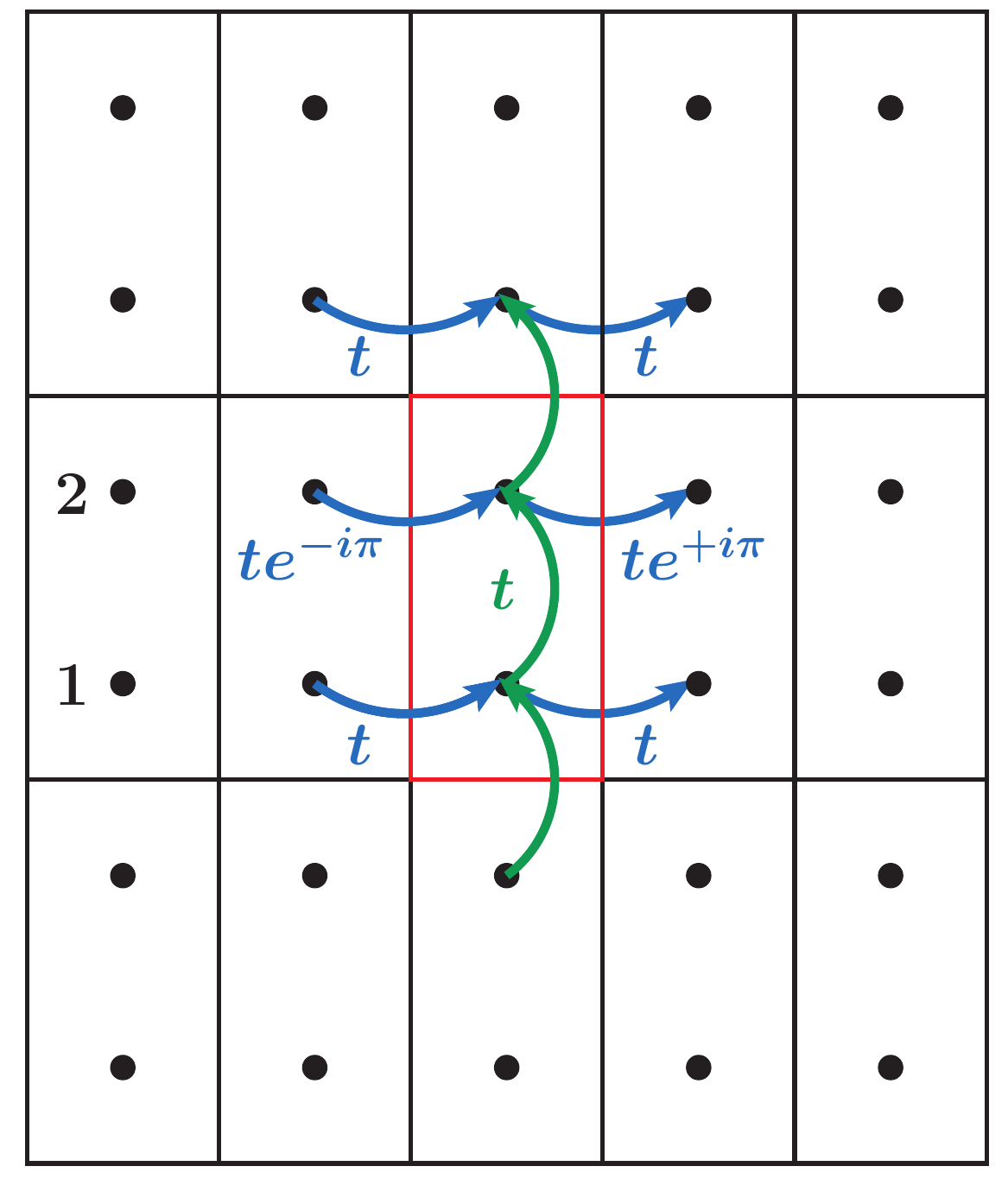}
		\caption{Division of lattice into magnetic unit cells (for $B\!=\!\Phi_0/2$).}
		\label{fig:magneticsupercell}
	\end{figure}
	
	Let us remark that the introduction of the Peierl's phase formally destroys the translational invariance of the system in $y$-direction. However, as the magnetic field is uniform, this should not affect any physical quantities for which translational invariance has to be recovered. Nevertheless, the formal breaking of translational symmetry poses technical problems for the actual calculations which usually exploit the existence of a well-defined (crystal)momentum quantum number in both $x$ and $y$ directions. Remarkably, the lattice obeys translational invariance with an increased period for a special set of values for the magnetic field $B$. In fact, if $B$ corresponds to a rational multiple of the magnetic flux quantum $\Phi_0$, i.e., $B\!=\!\frac{p}{q}\Phi_0$ (per unit cell $a^2\!\equiv\!1$), with $p,q\!\in\!\mathds{N}$ and gcd$(p,q)\!=\!1$, periodicity in $y$-direction is restored with a period of $q$ sites \cite{azbel1964energy} after which the phase factor becomes an integer multiple of $2\pi$. Hence, we can define a translational invariant lattice of so-called {\em magnetic unit cells} which contain $1$ site in $x$- and $q$ sites in $y$-direction (see Fig.\ref{fig:magneticsupercell}). The total flux which then penetrates such a magnetic unit cell is $p\Phi_0$.
		
	Introducing a lattice vector $\mathbf{\widetilde{R}}_i$ of the unit cell position allows us now to perform a Fourier transformation as follows: 
	
	\begin{equation}
	\label{equ:cfourier}
	c^{(\dagger)}_{i\sigma} = \sum_{\bm{k}} c^{(\dagger)}_{\bm{k}l\sigma} e^{\mp i\bm{k}\bm{\widetilde{R}}_i},
	\end{equation} 
	
	where the $\mathbf{k}$-integral has to be taken over the magnetic Brillouin zone \cite{zak1964magnetic} $k_x\!\in\!(-\pi,\pi)$, $k_y\!\in\!(-\frac{\pi}{q},\frac{\pi}{q})$ and the ``orbital'' index $l$ denotes the position inside the unit cell (see Fig.~\ref{fig:magneticsupercell}). The noninteracting part of the Hamiltonian (\ref{ham}) in this representation takes the form:
	
	\begin{widetext}
		\begin{equation} 			
		\begin{gathered}
		H_0 = \sum_{ll'\bm{k}\sigma}\varepsilon_{ll'}(\bm{k}) c^{\dagger}_{\bm{k}l\sigma}c_{\bm{k}l'\sigma}\\     
		\varepsilon_{ll'}(\bm{k})=
		\begin{mpmatrix}
		- 2 t \cos\left(k_x\right) & -t  & 0 & \ldots & t e^{iqk_y}\\
		-t & - 2 t \cos\left(k_x+ \frac{2 \pi p}{q}\right)& - t   & 0  &\ldots \\
		& & & \ddots&  & &\\
		t e^{-iqk_y}&0 &\ldots& -t &- 2 t \cos\left(k_x+ \frac{2 \pi p (q-1)}{q}\right),
		\end{mpmatrix}
		\end{gathered}
		\end{equation}
	\end{widetext}
	
\begin{figure}[t!]
		\centering
		\begin{subfigure}{1\linewidth}	
			\includegraphics[width=0.7\linewidth]{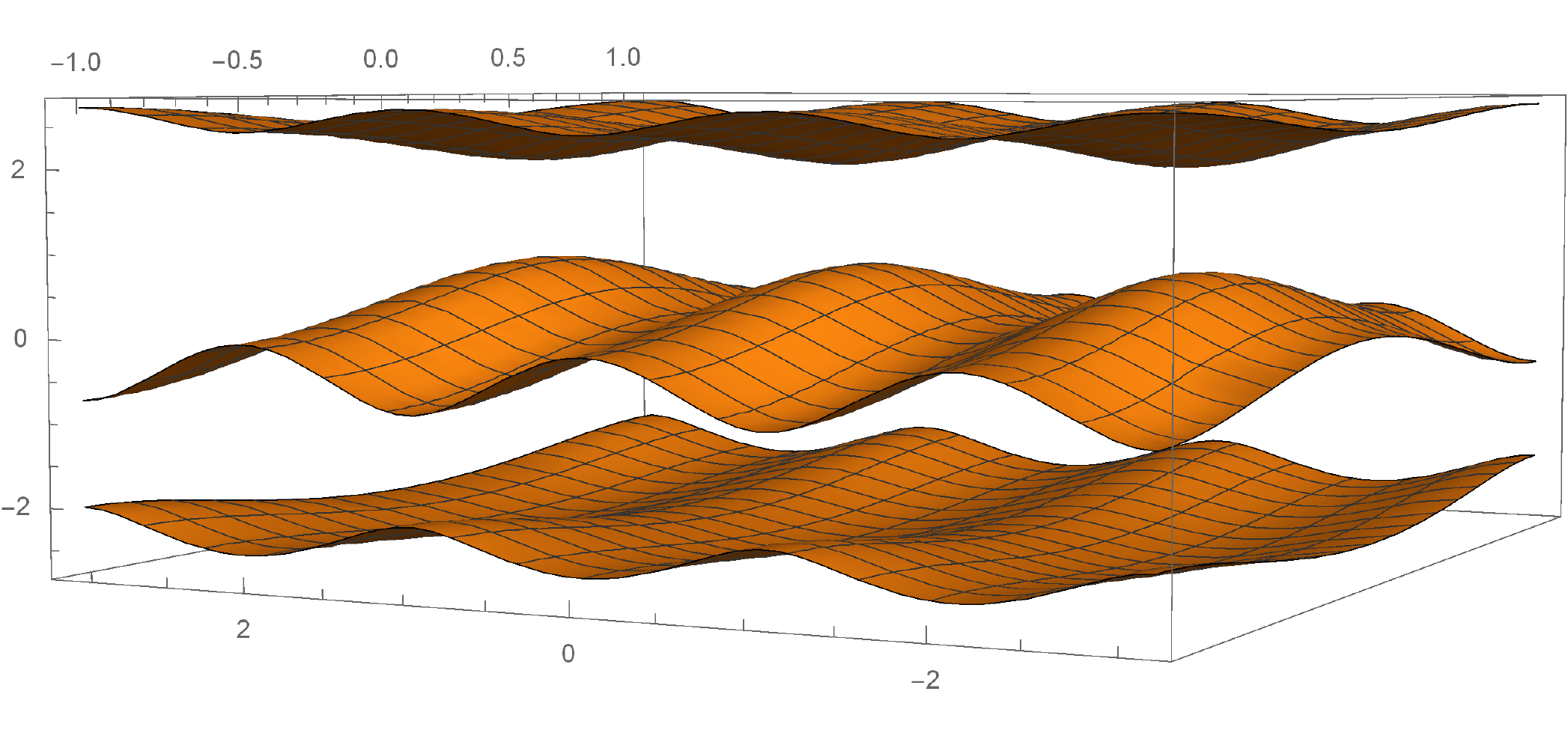}
		\end{subfigure}
		\begin{subfigure}{1\linewidth}	
			\includegraphics[width=0.8\linewidth]{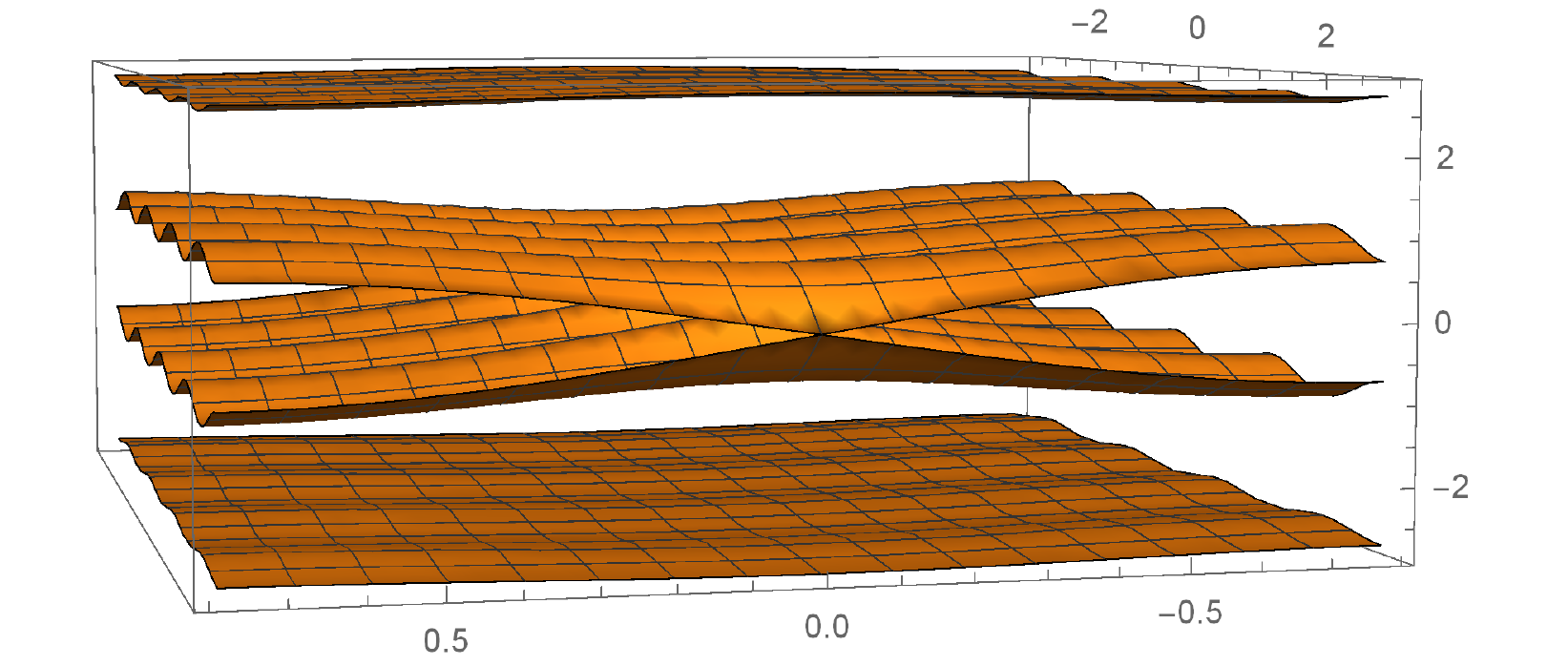}
		\end{subfigure}
		\caption{The dispersion relation of the Hofstadter model for $q=3$ and $q=4$.}
		\label{dispersion}
	\end{figure}

	The energy levels of $H_0$ can be now obtained by diagonalizing $\varepsilon_{ll'}(k)$ which is referred to as a Harper's equation \cite{harper1955single}. The resulting spectrum\footnote{Let us note that for magnetic fields which are irrational multiples of the flux quantum no band structure can be defined and the energy spectrum takes the form of a Cantor set.} as a function of $p/q$ is known as ``Hofstadter's butterfly'' \cite{hofstadter1976energy}. While $\varepsilon_{ll'}(k)$ can be diagonalized in general only numerically, some features of the system can be obtained analytically from the properties of the magnetic translation group \cite{wen1989winding}. The spectrum, for instance, is symmetric with respect to $\varepsilon=0$ and consists of $q$ bands periodic in $k_x$ and $k_y$ with the period $2\pi$ and $\frac{2\pi}{q}$, respectively. Moreover, for an even $q$, there are $q$ Dirac points at $\varepsilon=0$ \cite{kohmoto1989zero}. In Fig.~\ref{dispersion} the dispersion relations for $q\!=\!3$ and $q\!=\!4$ orbitals are shown. 
	
	\subsection{\label{top} The Hall conductivity at zero temperature}

	% While the correspondence between $\sigma_{xy}$ and the topological invariant $\mathcal{C}_1$ in Sec.~\ref{top} holds rigorously only at zero temperature and for non-interacting systems (or interacting systems with non-degenerate ground states and no closings of a gap), the many-body formalism in Sec.~\ref{DMFT} will allow us to calculate $\sigma_{xy}$ in the most general case for finite temperatures and arbitrary interactions.
	
%	\subsection{\label{top}Topological perspective}
	
%	$\bullet$ Kubo formula. The Integer Quantum Hall effect as a topological invariant at T=0, U=0.
	
	In this section we will discuss the Hall conductivity $\sigma_{xy}$ from the topological  perspective, which is applicable at zero temparature for both non-interacting and interacting systems. The only limitations are  a non-degenerate ground state and no changes in the gap structure. 
	
\begin{figure}[t]
		\centering
		\includegraphics[width=0.9\linewidth]{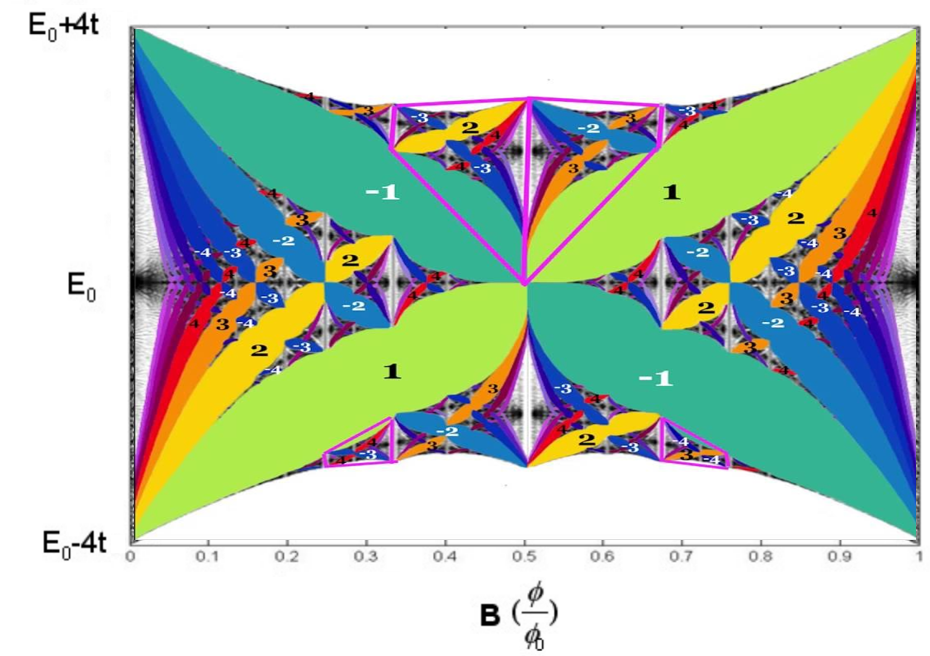}
		\caption{Gaps of the Hofstadter's butterfly colored correspondingly to their Hall conductivity \cite{satija2016butterfly} as a function $\varepsilon$ of $p/q$.}
		\label{cbut}
	\end{figure}

	In their seminal article, Thouless {\sl et al.} \cite{thouless1982quantized} demonstrated, that in the noninteracting case the Kubo formula \cite{kubo1957statistical} for the Hall Conductivity implies that its quantization can be explained by topological arguments. In fact, in this situation $\sigma_{xy}$ corresponds to the first Chern number \cite{simon1983holonomy} $\mathcal{C}_1$ of the $U(1)$ fiber bundle of the eigenvectors of the Hamiltonian on the torus defined by the Brillouin zone. More concretely, this topological invariant is an integral over the corresponding curvature for all filled bands which corresponds to an integer number if the chemical potential is located within a band gap, i.e., when each band is either contributing as a whole or not at all. In this way, an integer number can be assigned to each gap in the spectrum of the Hofstadter model which corresponds to the sum of all Chern numbers of the lower-lying bands and, hence, represents the Hall conductivity which is then associated to this gap. In Fig.~\ref{cbut}, the gaps of the Hofstadter model are colored according to these values of the Hall conductivity. Topological invariants are robust against continuous deformations, which nicely explains the phenomenal stability and robustness of the Hall plateaus observed in the famous experiments of von Klitzing \cite{Klitzing1980} and naturally raised a question about the limits of this topological protection.
	
	In this respect, a central question concerns the effect of interactions between the particles on the $\sigma_{xy}$ since electrons are typically subject to a (strong) Coulomb repulsion among each other. For the temperature $T\!=\!0$ the topological considerations can be extended to the case of interacting electrons \cite{niu1984quantised,avron1985quantization} by representing $\sigma_{xy}$ in terms of a {\em winding number} \cite{niu1984quantised} of the single-particle Green's functions:
	
	\begin{equation}
	\label{wind}
	\sigma_{xy} = C \epsilon_{ijl}\int d^3k\; Tr(G\partial_{k_i}G^{-1}G\partial_{k_j}G^{-1}G\partial_{k_l}G^{-1}),
	\end{equation}
	where $k_0\widehat{=}i\omega$ in order to obtain this symmetric form with respect to frequency and momentum variables and $C$ is a normalization constant. This formula can be derived under the assumption of a non-degenerate ground state \cite{ishikawa1986magnetic}  (which is not true in the case of the Fractional Quantum Hall Effect). Mathematically, the Green's function $G(k)$ in the framework of Eq.~(\ref{wind}) can be interpreted as a map from the momentum space to the $GL(N,\mathbb{C})$. The expression on the r.h.s. of this equation is referred to as winding number since, loosely speaking, it counts the number of times the momentum space is wrapped around the target space $GL(N,\mathbb{C})$. Note that a general rigorous mathematical proof of the topological invariance of (\ref{wind}) is not that simple, especially taking into account that the domain $T^2\times \mathbb{R}$ is not compact from the first glance (a schematic justification of the fact is given in  \cite{ishikawa1986magnetic,qi2008topological}). In the noninteracting limit, Eq.~(\ref{wind}) indeed reduces to the above discussed first Chern number \cite{budich2013fluctuation}. With that in mind, let us now analyze, what happens upon switching on the Hubbard interaction $U$. 
	
	It was shown that as long as $G(\omega,k)$ has no singularities at $\omega=0$ one can smoothly deform it and obtain a Green function of a non-interacting model with the same value of $\sigma_{xy}$ \cite{wang2012simplified}. So the effect of non-zero $U$ corresponds to a shift and/or smooth deformation of the noninteracting bandstructure w.r.t. the related non-interacting model. This picture holds, in particular, if no symmetry breaking occurs at low temperatures and the system is a Fermi Liquid (FL) at $T\!=\!0$. In this situation, the above mentioned requirements are fulfilled and the gap structure as well as the associated topological invariant integer values for $\sigma_{xy}$ remain stable.
	
	A further simplification occurs if one assumes the locality of the self-energy function $\Sigma(i\omega_n)$, which is the case for the DMFT analysis presented below. In this situation, only the positions and the sizes of the gaps can be changed through an energy dependent renormalization of the chemical potential by the real part of the self-energy Re$\Sigma(\omega=0,\mu)$ (where $\mu$ denotes the chemical potential). The only possibility to qualitatively modify the physical picture in this case is the emergence of a zero in the Green's function as it is observed for the onset of a Mott insulating phase.
	
	It has to be stressed, that the above considerations are only valid at $T\!=0\!$ where for a FL the lifetime of the corresponding quasi particles is infinite. In the following, we are however interested in the combined effect of interaction and finite temperature on $\sigma_{xy}$. In this situation Eq.~(\ref{wind}) is not applicable any more since it requires {\em continuous} Matsubara frequencies which are well defined only at $T\!=\!0$. %In the following section we will, hence, present an expression for $\sigma_{xy}$ which is applicable in the most general situations.   

	\subsection{\label{DMFT} The Hall conductivity at finite temperatures}

	In order to treat the interacting system Eq.~(\ref{ham}) at finite temperature we apply the DMFT approximation \cite{georges1996dynamical}. Within this approach, the actual lattice is replaced by a single interacting site embedded into a self-consistent bath. This corresponds to an Anderson impurity model (AIM) whose non-interacting part is defined by the requirement that the one-particle Green's function of the AIM is identical to the local (i.e., $\mathbf{k}$-integrated) one-particle Green's function of the lattice:
	
	\begin{equation} 
	G^{\text{AIM}}(i\omega_n) = \sum_{\bm{k}} [(i\omega_n + \mu -\Sigma(i\omega_n))\delta_{ll'} - \varepsilon_{ll'}(\bm{k})]^{-1}\Bigr\rvert_{l=l'}
	\label{consist}
	\end{equation} 
	where $\omega_n\!=\!(2n\!+\!1)\frac{\pi}{\beta},n\!\in\!\mathds{Z}$, denotes a fermionic Matsubara frequency and $\Sigma(i\omega_n)$ is the local  self-energy of the auxiliary AIM. As in Eq.~(\ref{equ:cfourier}), $\sum_{\mathbf{k}}$ denotes the momentum integration over the magnetic Brillouin zone. Let us emphasize that, although formally we are concerned with a multi-orbital systems, $G^{\text{AIM}}(i\omega_n)$ and $\Sigma(i\omega_n)$ are scalars. From a general physical perspective, this simplification is obvious since the magnetic field is uniform and, hence, the auxiliary AIM has to be the same for all lattice sites which reduces the multi- to a single-orbital DMFT problem (i.e., $G^{\text{AIM}}_{ll'}\!\equiv\!G^{\text{AIM}}\delta_{ll'}$, and the same for $\Sigma$). A more formal argument for this simplifications follows from the gauge invariance of the system: While the entire lattice Green's function appears to be different for different choices of the gauge\footnote{Note, that for different gauges the definition of the magnetic unit cell will differ. For instance, in the alternative gauge $\mathbf{\widetilde{A}}(\mathbf{r})\!=\!-B(y+1,0,0)$ the lattice sites in the unit cell are shifted by $1$ ($l\!\rightarrow\!l\!+\!1$)}, its local part does not vary under a gauge transformation. This is only possible if the local part of Green's function is the same for all lattice sites within the magnetic cluster.  Therefore the AIM for different sites within the cluster are identical.

% In the present situation, the different orbitals correspond to lattice sites within the magnetic unit cells which requires that $\Sigma_{ll'}(i\omega_n)\!\propto\!\delta_{ll'}$ since the DMFT self-energy is local. Moreover, all the lattice sites withing the magnetic unit cell are physically equivalent because the magnetic field is uniform which implies that $\Sigma_{ll'}(i\omega_n)\!=\!\Sigma(\omega)\delta_{ll'}$. Finally, the self-consistency relation~(\ref{consist}) is obviously valid only for $l\!=\!l'$ which reduces the multi-orbital to a standard single-orbital DMFT problem. In this respect, let us point out that the actual construction of the magnetic unit cell depends on the chosen gauge (see also Appendix~\ref{app:novertcorr}) which also requires an orbital independent self-energy in order to preserve the gauge invariance of the system.

	%$\bullet$ General formula for $\sigma_{xy}$ at $T\neq0$ and $U\neq0$

	In order to calculate the Hall conductivity for $T>0$ and $U>0$, we start from the most general form of the linear response Kubo formula \cite{kubo1957statistical} for the optical conductivity in the dipole approximation $\mathbf{q}=0$ which is given by
	\begin{align}
	\label{equ:sigmageneral}
	\text{Re}\sigma^{\alpha\beta}(\nu,\mathbf{q}=0)&=-\frac{\text{Im}\chi_{\text{jj}}^{\alpha\beta}(\nu,\mathbf{q}=0)}{\nu}\nonumber\\&= -\frac{\text{Im}\chi_{\text{jj}}^{\alpha\beta}(i\nu_m\rightarrow\nu+i0^+,\mathbf{q}=0)}{\nu},
	\end{align}
	where $\nu$ denotes a frequency on the real frequency axis while $i\nu_m\!=\!\frac{2m\pi}{\beta}$, $m\in\mathds{Z}$, is a bosonic Matsubara frequency and $i\nu_m\rightarrow\nu+i0^+$ indicates the analytical continuation of the function from the upper complex half plane to the real axis. $\chi_{\text{jj}}^{\alpha\beta}(i\nu_m,\mathbf{q}=0)$ is the (paramagnetic) current-current correlation function which reads
	\begin{align}
	\label{equ:currentcurrentgeneral}
	\chi^{\alpha\beta}&(i\nu_m,\mathbf{q}=0)=-\frac{e^2}{\hbar} \sum_{\mathbf{k}\mathbf{k'}}\sum_{l_1l_2l_3l_4}\sum_{\sigma\sigma'}v_{\mathbf{k},l_1l_2}^{\alpha}v_{\mathbf{k'},l_3l_4}^{\beta}\nonumber\\&\times \int_0^\beta e^{i\nu_m\tau}\left\langle\mathcal{T}c^\dagger_{\mathbf{k}l_1\sigma}(\tau)c_{\mathbf{k}l_2\sigma}(\tau)c^\dagger_{\mathbf{k'}l_3\sigma'}(0)c_{\mathbf{k'}l_4\sigma'}(0)\right\rangle,
	\end{align} 
	where $\mathcal{T}$ denotes the time-ordering operator, $c^{(\dagger)}_{\mathbf{k}l\sigma}$ are the (creation) annihilation operators in momentum representation and $\sum_{\mathbf{k}}$ corresponds to a normalized momentum integration over the magnetic Brillouin zone. The Fermi velocities are given by:
	\begin{align}
	\label{equ:defvolicity}
	v_{\mathbf{k},ll'}^{\alpha}=\frac{\partial\varepsilon_{\mathbf{k}}^{ll'}}{\partial k_\alpha}-i(\rho_l^\alpha-\rho_{l'}^\alpha)\varepsilon_{\mathbf{k}}^{ll'},
	\end{align}
	where $\rho_l^\alpha$ denotes the relative position of the site $l$ inside the magnetic unit cell. The matrix element $\langle\ldots\rangle$ in Eq.~(\ref{equ:currentcurrentgeneral}) corresponds to the full two-particle Green's function of the system. It can be decomposed \cite{rohringer2012local} into the bubble term\footnote{The second possible bubble contraction \unexpanded{$\langle\mathcal{T}c^\dagger_{\mathbf{k}l_1\sigma}(\tau)c_{\mathbf{k'}l_2\sigma}(\tau)\rangle\langle\mathcal{T}c^\dagger_{\mathbf{k'}l_3\sigma'}(0)c_{\mathbf{k}l_4\sigma}(0)\rangle$} is independent of $\tau$ and, hence, its Fourier transform to imaginary frequencies $i\nu_m$ yields a term proportional to $\delta_{\nu_m0}$ which does not contribute when the analytic continuation $i\nu_m\rightarrow\nu+i0$ if performed.}:
	$-\langle\mathcal{T}c^\dagger_{\mathbf{k}l_1\sigma}(\tau)c_{\mathbf{k'}l_4\sigma'}(0)\rangle\langle\mathcal{T}c^\dagger_{\mathbf{k'}l_3\sigma'}(0)c_{\mathbf{k}l_2\sigma}(\tau)\rangle=G_{l_1l_4}(\tau,\mathbf{k})G_{l_3l_2}(-\tau,\mathbf{k})\delta_{\sigma\sigma'}\delta_{\mathbf{k}\mathbf{k'}}
	 $ and a contribution containing vertex corrections. 
	 
	Let us present a generic argument why vertex corrections are absent for the system under consideration. When calculating the susceptibility $\frac{d (i \omega +\mu-\varepsilon-\Sigma)^{-1}}{d {\cal E}}$ to the external electric field ${\cal E}$, the vertex corrections appear because the self-energy part  $\Sigma$ varies with $\varepsilon$. Within DMFT, the self-energy originates from the self-consistent AIM defined by Eq.~(\ref{consist}). Therefore the vertex correction vanishes if the AIM does not vary under the (infinitesimal) external electric field. Let us introduce a uniform d.c. field along the $x$-direction by a (time dependent) Peierls phase in the corresponding hopping terms: $t_{x, x+1}\to t_{x, x+1} e^{-i {\cal{E}} s}$ and conjugate for $t_{x+1, x}$, where $s$ denotes the time. The system remains periodic in $x$-direction with the initial lattice constant (remember that, in the chosen gauge, the magnetic unit cell extends along the $y$ direction). In momentum space,  the effect of the applied electric field is a ``drift'' of the energy levels of the noninteracting Hamiltonian $H_0$: Its part associated with the hopping in $x$-direction equals $H_0^x=\sum_{k_x} 2 t \cos (k_x-{\cal E} s)$. Now we observe that (\ref{consist}) contains a sum over all possible $k_x$. Clearly, the time-dependent offset ${\cal E}$ does not affect the value of the sum. Therefore $\cal{E}$ drops out from the DMFT self-consistency condition, and the AIM is independent of it. This proofs that vertex corrections drop out  in the calculation of $\sigma^{xy}$ within the single-site DMFT. In principle, these considerations apply also for a time-dependent field. As far as we know, such a generic argument has not been presented so far. Without magnetic field, this fact is known as a result of the antisymmetry of the Fermi velocity under $\mathbf{k}\!\rightarrow\!-\mathbf{k}$. Appendix ~\ref{app:novertcorr} generalizes this more explicit proof to our case with a magnetic field. 
	
	With the DMFT simplification the actual expression for the current-current correlation function reads
	
	\begin{align}
	\label{equ:hallDMFTsimplification}
	\chi^{\alpha\beta}(\nu,\mathbf{q}=0)=\frac{2e^2}{\hbar}\frac{1}{\beta}&\sum_{\omega_n}\sum_{\mathbf{k}}\sum_{l_1l_2l_3l_4}v_{\mathbf{k},l_1l_2}^\alpha G_{l_2l_3}(i\omega_n,\mathbf{k})\nonumber\\&\times v_{\mathbf{k},l_3l_4}^\beta G_{l_4l_1}(i\omega_n+i\nu_m,\mathbf{k}),
	\end{align}
	where the sum over $l_1\ldots l_4$ corresponds to the trace in orbital space. This trace is invariant under a transformation to an orbital basis in which $G_{ll'}(i\omega_n,\mathbf{k})$ [and $\varepsilon_{ll'}(\mathbf{k})$] are diagonal. Moreover, we can represent $G_{ll'}(i\omega_n,\mathbf{k})$ by its spectral representation which yields

	\begin{align}
	\label{equ:calcoptDMFT}
	\chi^{\alpha\beta}&(\nu,\mathbf{q}=0)=-\frac{2e^2}{\hbar}\sum_{\mathbf{k}}\sum_{l_1l_2}\int d\omega\int d\omega'\nonumber\\\times&\frac{f(\omega')-f(\omega)}{\omega-\omega'+\nu+i\delta}\left[\bar{v}_{\mathbf{k},l_1l_2}^\alpha A_{l_2}(\omega,\mathbf{k})\bar{v}_{\mathbf{k},l_2l_1}^\beta A_{l_1}(\omega',\mathbf{k})\right],
	\end{align}
	where $f(\omega)=(1+e^{\beta\omega})^{-1}$ denotes the Fermi function. The spectral functions $A_{l}(\omega,\mathbf{k})=-\frac{1}{\pi}\text{Im}\bar{G}_{ll}(i\omega_n\rightarrow\omega+i\delta,\mathbf{k})$ are defined in the standard way whereas $\bar{G}_{ll'}(i\nu_n,\mathbf{k})\!=\!\bar{G}_{ll}(i\nu_n,\mathbf{k})\delta_{ll'}$ and $\bar{v}^{\alpha/\beta}_{\mathbf{k},ll'}$ correspond to the Green's function and the Fermi velocity in the transformed orbital space, respectively (where the Green's functions are diagonal). In order to calculate the optical conductivity $\sigma^{\alpha\beta}(\nu)$ we have to consider the imaginary part of $\chi^{\alpha\beta}(\nu,\mathbf{q}=0)$. To this end we note that $1/(\omega-\omega'+\nu+i\delta)=\text{P}\frac{1}{\omega-\omega'+\nu}-i\pi\delta(\omega-\omega'+\nu)$. For the standard case of the direct conductivity ($\alpha=\beta$) in systems without magnetic field the second (trace) term of Eq.~(\ref{equ:calcoptDMFT}) is typically purely real and, hence, all contributions originate from $i\pi\delta(\omega-\omega'+\nu)$ which -- for small values of $\nu$ -- is mainly governed by the spectral weight at the Fermi level. In our case, however, such contribution vanishes for symmetry reasons.
	
	Instead, for the Hall conductivity ($\alpha=x$, $\beta=y$) the orbital trace in Eq.~(\ref{equ:calcoptDMFT}) acquires an imaginary part and, hence, the principal value $\text{P}\frac{1}{\omega-\omega'+\nu}=\frac{\omega-\omega'+\nu}{(\omega-\omega'+\nu)^2+\delta^2}$ yields a finite contribution to  the Hall conductivity which for $\nu\rightarrow 0$ becomes
	\begin{align}
	\label{equ:principalvaluecontribution}
	\sigma^{xy}&(\nu=0)=-\frac{2e^2}{\hbar}\int d\omega \int d\omega' \frac{f(\omega)-f(\omega')}{(\omega-\omega')^2}\nonumber\\\times &\sum_{\mathbf{k}}\sum_{l_1l_2}\bar{v}_{\mathbf{k},l_1l_2}^x A_{l_2}(\omega,\mathbf{k})\bar{v}_{\mathbf{k},l_2l_1}^y A_{l_1}(\omega',\mathbf{k}).
	\end{align}
	This term acquires contributions mainly from energies away from the Fermi level and, hence, will be finite also if the system is gaped (in contrast to the $\delta$-like term discussed above whose $\nu=0$ contribution would vanish in this case). A straightforward calculation shows that for $U\rightarrow0$ and $T\rightarrow 0$ Eq.~(\ref{equ:principalvaluecontribution}) indeed reduces (up to a prefactor) to the topological invariant, i.e., the first Chern number $\mathcal{C}_1$ discussed in the previous section. Hence, Eq.~(\ref{equ:principalvaluecontribution}) allows us to study how $\sigma^{xy}(\nu=0)$ is modified by a combination of finite temperatures and interactions and, in particular, how deviations from its topologically determined integer values emerge.

	\begin{figure}[t]
	\centering
	\begin{subfigure}{1\linewidth}	
		\includegraphics[width=0.9\linewidth]{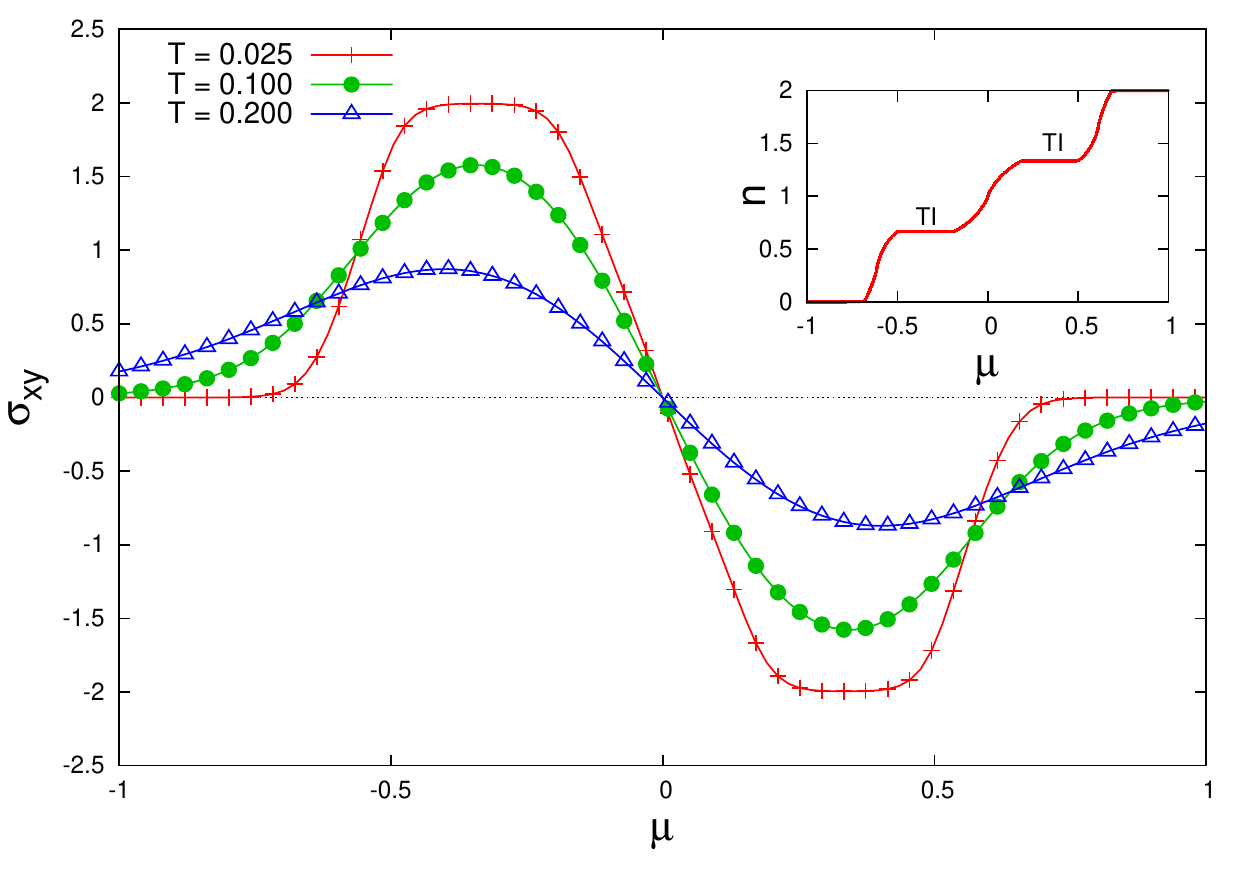}
	\end{subfigure}
	\begin{subfigure}{1\linewidth}	
		\includegraphics[width=0.9\linewidth]{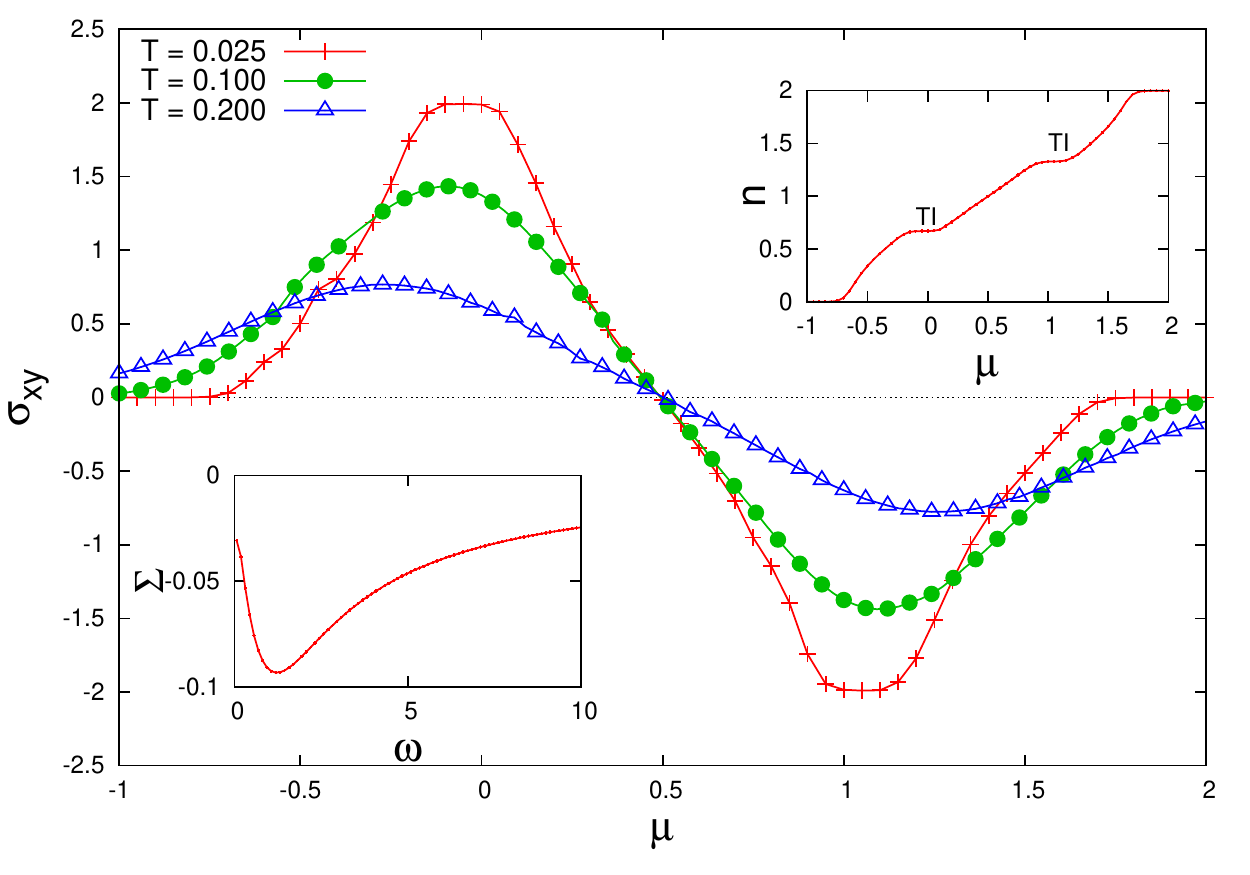}
	\end{subfigure}
	\begin{subfigure}{1\linewidth}	
		\includegraphics[width=0.9\linewidth]{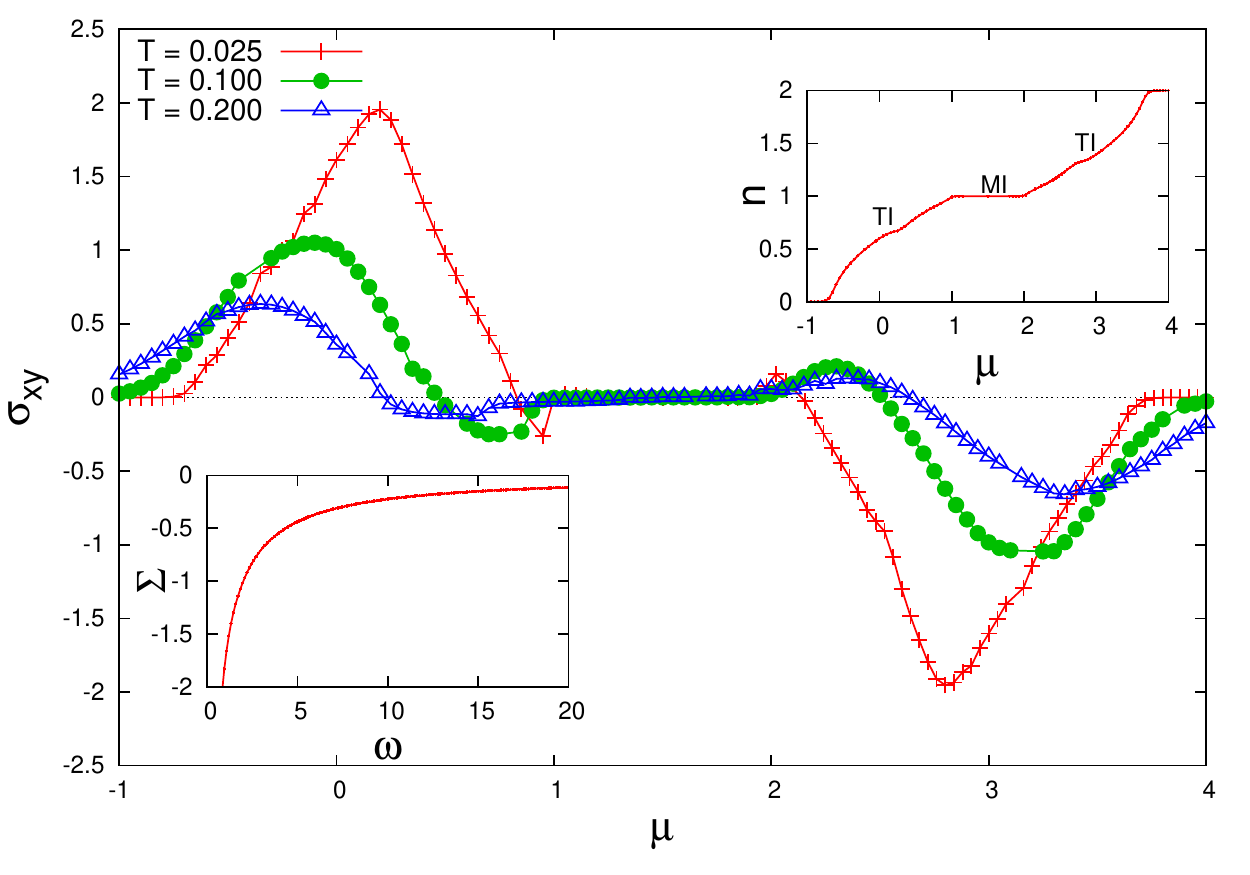}
	\end{subfigure}
	\caption{Hall conductivity vs. chemical potential for different values of $U$ and $T$. The insets in the right upper corner show the electron density vs. the chemical potential at $T = 0.25$ . For the two interacting cases ($U\!\ne\!0$) the imaginary part of self-energy at half-filling are also presented for $T = 0.025$ in the left bottom insets.}
	\label{fig:SigmaMu}
	\end{figure}	
	
	\section{\label{results}Results}
	%$\bullet$ Description of numerical procedures: ED, Pade, $\sigma_{xy}$
	
	We have calculated the dc Hall conductivity $\sigma^{xy}(\nu=0)\!\equiv\!\sigma^{xy}$ within the DMFT approximation for different values of the Hubbard interaction $U$, the temperature $T$ and the particle density $n$ (or, correspondingly, the chemical potential $\mu$) by means of Eq.~(\ref{equ:principalvaluecontribution}). For the calculation of the DMFT self-energy $\Sigma(i\omega_n)$ on the imaginary frequency axis we have adopted an exact diagonalization (ED) algorithm with $5$ bath sites whose stability has been checked w.r.t. larger bath sizes. In principle, within ED, results for $\Sigma$ could be directly obtained also on the real frequency axis. However, in this case the corresponding spectral function consists just in a number of $\delta$-peaks which reflects the artificial nature of the finite bath. In order to obtain more physical (i.e., continuous) expressions for $\Sigma(i\omega_n\rightarrow\omega+i\delta)$ we have performed an analytic continuation of the Matsubara data to the real frequency axis by means of a Pade fit which, in turn, have been then used for the calculation of $A_{l}(\omega,\mathbf{k})$ and $\sigma^{xy}$ in Eq.~(\ref{equ:calcoptDMFT}). In the following, we present our numerical results for $\sigma^{xy}$ in units of $\frac{e^2}{2\pi \hbar}$. The energy scale is set by $t=0.25$.

	\subsection{ The Hall conductivity for different values of U}

\begin{figure}[t]
		\centering
		\includegraphics[width=1\linewidth]{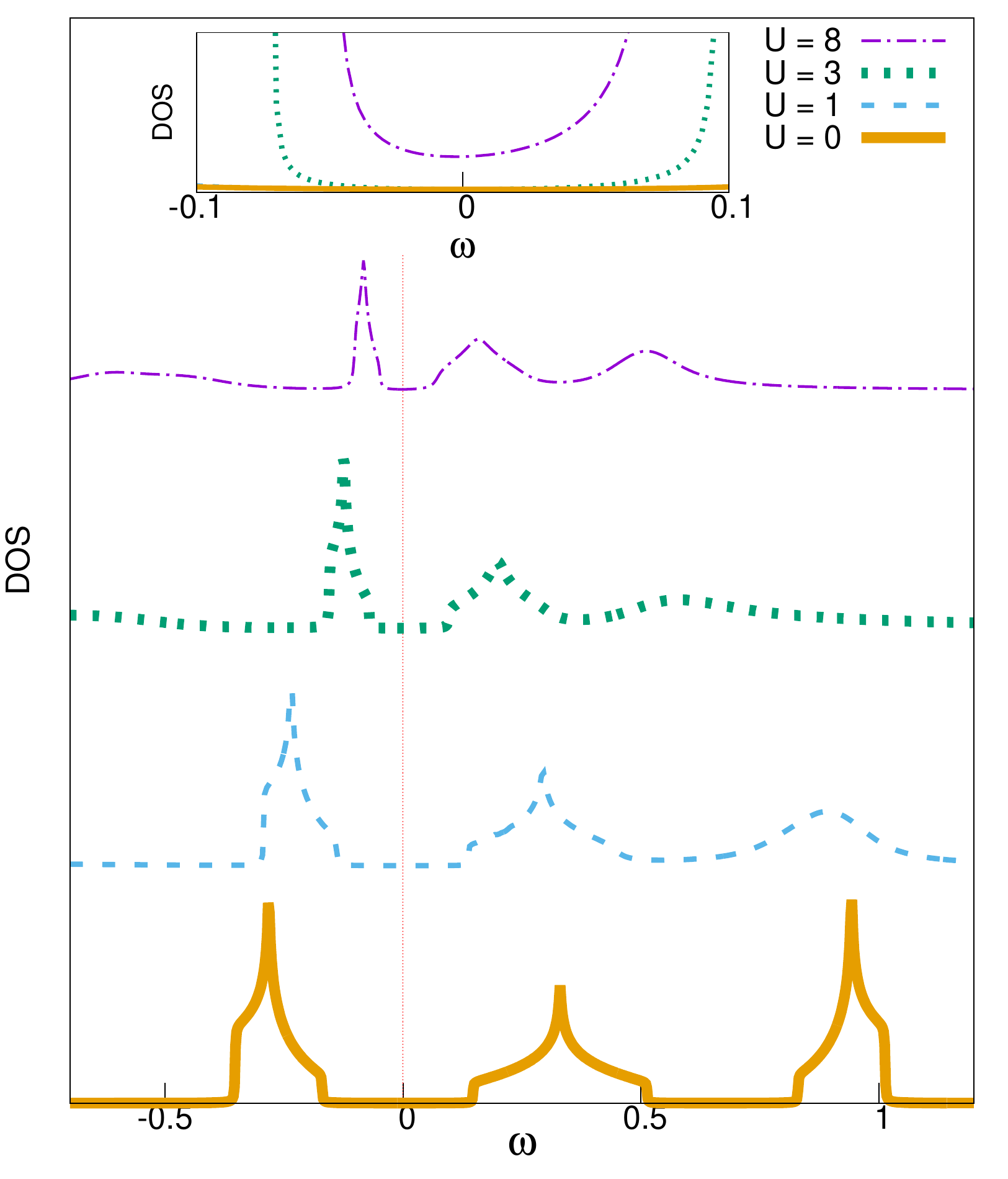}
		\caption{Density of states at different values of $U$ for $T=0.025$. Here, we fixed the value $\mu - \text{Re}\Sigma(\omega = 0)$ in order to keep the effective Fermi level close to the middle of the first gap.}
		\label{fig:DOS}
	\end{figure}

	We have selected a rather large magnetic field strength $B\!=\!\frac{1}{3}\Phi_0$ for which the system is described by $q\!=\!3$ bands and, hence, exhibits two (band) gaps (orange curve in Fig.~\ref{fig:DOS}) which -- at $T=0$ -- correspond to the quantized Hall conductivities (and first Chern number $\mathcal{C}_1$) $-2$ and $2$ (where a factor of of $2$ originates from the spin). In Fig.~\ref{fig:SigmaMu} the results for $\sigma^{xy}$ for three different temperatures $T$ and values of the interaction strength $U$ are presented as a function of the chemical potential $\mu$. %Even though, the chemical potential from the experimental point of view is less physical quantity then average density of electrons, the dependency of $\mu$ reveals information about gaps in the system. Since these play the key role in effects we are interested in, we have preferred $\mu$ to the average density. However, one should keep in mind that the integer plateaus presented on the dependencies are somewhat slightly different then usual famous Hall plateaus.
	
	For the non-interacting case $U=0$ (upper panel) and the lowest temperature $T=0.025$, we can clearly identify the two plateaus of $\sigma^{xy}(\nu=0)$ at $2$ and $-2$ which correspond to the two gaps in the system. Upon increasing the temperature, we find a suppression of $\sigma^{xy}$ which originates solely from the broadening due to the Fermi functions in Eq.~(\ref{equ:principalvaluecontribution}) since the spectral functions are temperature independent in the non-interacting case.
		\begin{figure}[t]
		\centering
		\begin{subfigure}{1\linewidth}	
			\includegraphics[width=0.9\linewidth]{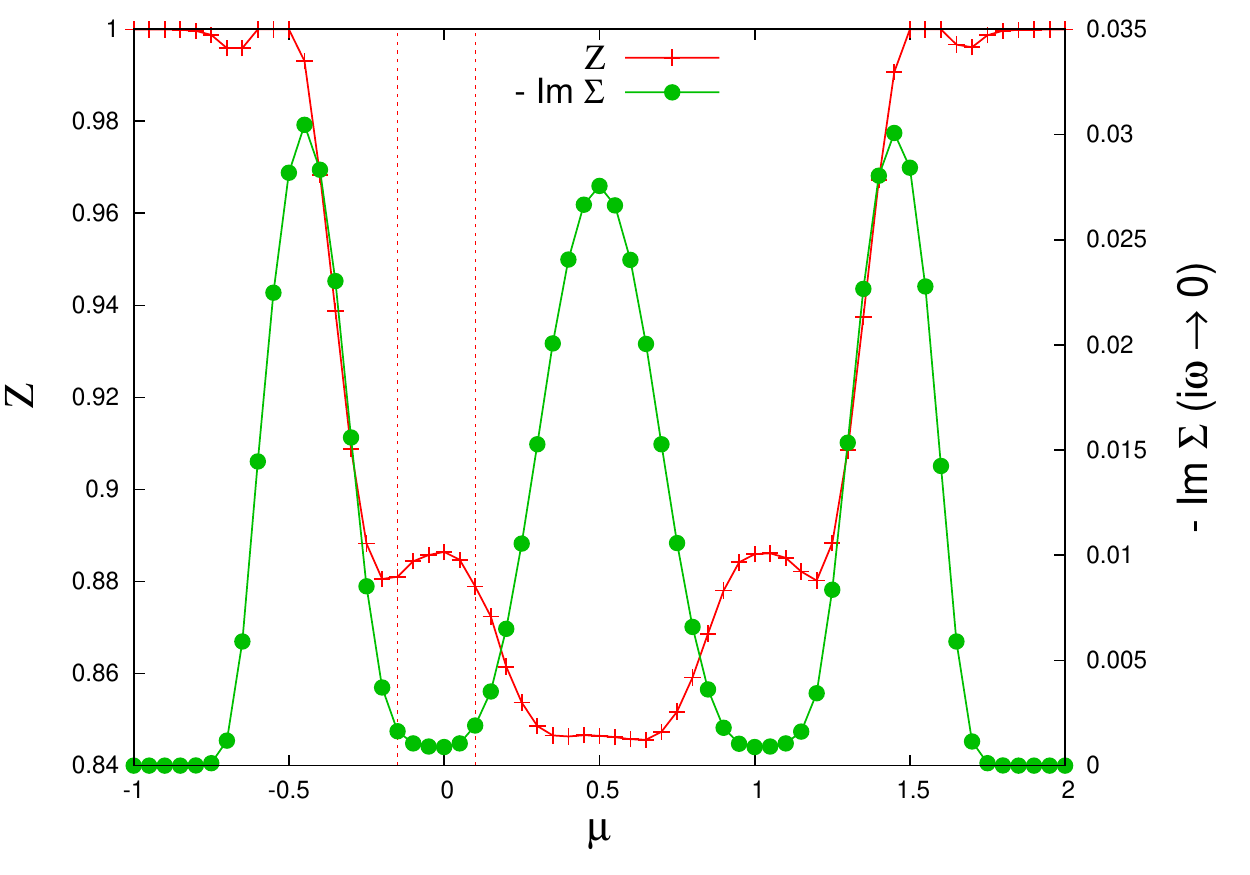}
		\end{subfigure}
		\begin{subfigure}{1\linewidth}	
			\includegraphics[width=0.9\linewidth]{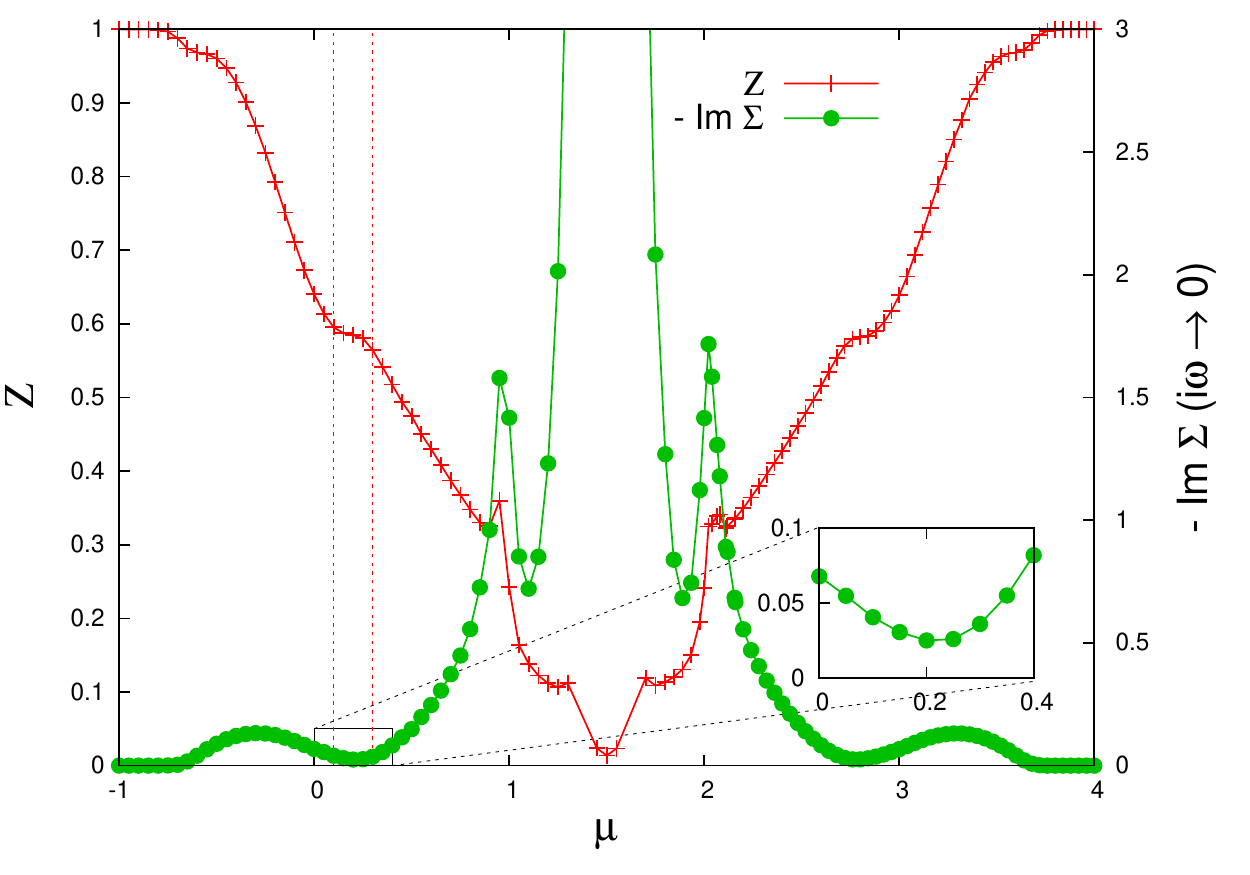}
		\end{subfigure}	
		\caption{Quasiparticle weight $Z\!=\!(1\!-\!\partial\text{Re}\Sigma/\partial\omega\lvert_{\omega\!=\!0})^{-1}$ (left $y$-axis) and inverse quasiparticle lifetime $\gamma\!=\!-\text{Im}\Sigma(\omega\!=\!0)$ (right y-axis) as a function of chemical potential $\mu$, for $U\!=\!1$ (upper panel) and $U\!=\!3) $ (lower panel), respectively. Vertical lines indicate the location of the first (band) gap. The inset shows $\gamma$ for $U\!=\!3$ inside the first band gap on a smaller scale.}
		\label{fig:Quas}
	\end{figure}
		%$\bullet$ DOS plot to illustrate these features:

	%Compare the gap widths.

	For the weak-to-intermediate coupling $U=1.0$ (middle panel) one observes a narrowing of the plateaus at $2$ and $-2$ which corresponds one-to-one to a correlation-driven narrowing of the two gaps in the spectral function in Fig.~\ref{fig:DOS}. For this value of $U$, the system is still metallic at half-filling as it can be seen from the corresponding self-energy on the Matsubara axis in the lower left inset of the figure. Hence, for this value of the interaction, correlation effects do not qualitatively change the physical picture but just lead to a renormalization w.r.t. the noninteracting case. This is also illustrated in the upper panel of the Fig. ~\ref{fig:Quas} where the value of the quasiparticle renormalisation factor $Z$ close to $1$ ($Z\!\sim\!0.9$) in the middle of the first gap indicates the coherence of the system and the rather small narrowing of the gap (as in the case of Fermi liquid $\Delta = Z \Delta^0$ where $\Delta$ and $\Delta^0$ are the gap width for the interacting and noninteracting system, respectively). This coherence is also reflected in the very small value of the quasiparticle scattering rate (inverse lifetime) $\gamma$ inside the first gap which is of the order $10^{-3}$.   
	
	Finally, at the largest value of $U=3.0$ at $T=0.025$ the plateaus which indicate the integer quantum Hall effect have almost disappeared. Interestingly, at this value of $U$ there is still a sizable gap in the spectral function. This suggests that the suppression of $\sigma^{xy}$ is not only due to the redistribution of spectral weight but also strongly influenced by the incoherence in the system. In fact, the quasiparticle weight in the center of the first gap (see lower panel of Fig.~\ref{fig:Quas}) is already reduced to $Z\!\sim\!0.6$ and the scattering rate $\gamma$ is an order of magnitude larger than for $U\!=\!1$ (see inset in the lower panel of Fig.~\ref{fig:Quas}). In addition, the Hall conductivity exhibits a plateau around half-filling ($\mu=1.5$) which corresponds to the incompressibility of the system  and indicates that -- at half-filling -- the system is already in the Mott phase. This is also confirmed by the self-energy at half-filling (see lower left inset in Fig~\ref{fig:SigmaMu}) which exhibit already an insulating behavior. 
		
	Another interesting observation for $U=3$ is the unexpected sign change of the Hall conductivity right before (and, correspondingly) after half-filling. This corresponds to a change in the type of charge-carriers from electrons to holes even before half-filling is reached (where such a sign change is obvious). In  \cite{rojo1993sign} it has been demonstrated that such a sign-change can be attributed to an anomalous behavior of the kinetic energy as a function of the density. Moreover, such a behavior is also consistent with the classical intuition about the slightly hole-doped Hubbard model, where almost all sites are single-occupied and only the holes can move. Let us, however, point out that this effect is reduced upon decreasing the temperature \cite{assaad1995hall} and the sign of the hole-doped system becomes again electron-like since according to Fermi liquid theory the scattering rate scales as $\sim T^2$ and the system becomes metallic again (for $T\!\rightarrow\!0$). In our case, such low temperatures can hardly be reached due to the limitations of ED. Nevertheless, one can observe the overall tendency in the lowest panel of Fig.~\ref{fig:SigmaMu}, where the interval of the anomalous sign of the Hall conductivity becomes smaller with decreasing temperatures.  
	
	\begin{figure}[t]
		\centering
		\includegraphics[width=1\linewidth]{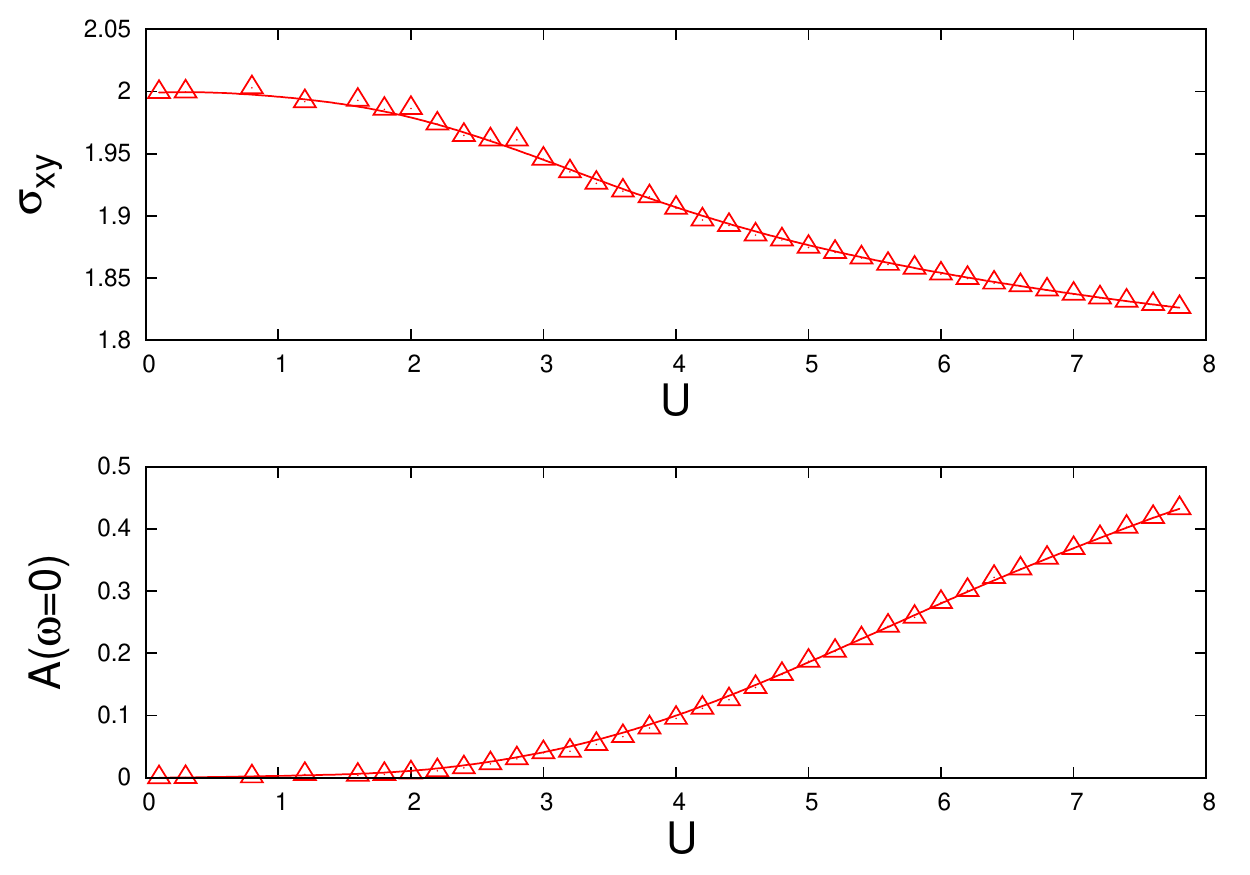}
		\caption{Dependence of $\sigma_{xy}$ and $A(\omega=0)$ on the interaction strength $U$ at $T=0.025$. }
		\label{fig:hallvsU}
	\end{figure}
	
	For the $U$ values considered so far, we have observed a reduction of the size of the integer quantum Hall plateaus due to correlation effects. It is, however, still unclear whether correlations can lead also to deviations from these integer values. This question is addressed in the upper panel of Fig.~\ref{fig:hallvsU} where the Hall conductivity is shown as a function of $U$ fixing the value of $\mu$ to the corresponding maximum of $\sigma^{xy}$ (within the gap). This corresponds to fixing the value $\mu - \text{Re}\Sigma(\omega = 0)$ so that the Fermi level remains in the center of the first gap (or what remains from this gap). On the suitably chosen scale, we indeed observe a tiny but clearly visible decrease of $\sigma^{xy}$ upon increasing $U$. This reduction can be attributed to a correlation induced filling of the gap which can be observed by a related increase of $A(\omega=0)$ as function of $U$ shown in the lower panel of Fig.~\ref{fig:hallvsU}. In fact, correlation effects begin to fill the gap starting from a value of $U\sim 2$ which exactly coincides with the interaction value at which $\sigma^{xy}$ begins to decrease, indicating a breakdown of the integer quantum Hall regime (which was still present at $U=1.0$ in Fig.~\ref{fig:SigmaMu}).

	%In order to obtain more insights into the correspondence between the filling of the gap in the spectral function and the decrease of $\sigma^{xy}$ we have plotted both quantities in the lower and upper panels of Fig.~\ref{fig:hallvsU}, respectively. For these calculations we have fixed the value $\mu - \text{Re}\Sigma(\omega = 0)$ in order to keep the Fermi level in the center of the first gap (or what is left over from this gap). We observe, that very small spectral weight (due to temperature) within the gap is almost independent of $U$ up to a value of $U\sim 2.0$ where the interaction between the particles starts to fill this gap (corresponding to an increase of $A(\omega=0)$). Correspondingly, the value of $\sigma^{xy}(\nu=0)$ starts to decrease at exactly this value of $U$ which indicates a breakdown of the integer quantum Hall regime (which was still visible at $U=1.0$ in Fig.~\ref{SigmaMu}).
	
	From a more general perspective we can argue, that the onset of correlations due to interactions between the particles suppresses the Hall conductivity $\sigma^{xy}$. The very same effect is typically observed for the normal dc conductivity $\sigma^{xx}$ in a metallic system, however, for the opposite reason. While for $\sigma^{xy}$ the suppression occurs due to the correlation induced transfer of spectral weight {\em to} the Fermi level at which we would otherwise have a perfect gap, $\sigma^{xx}$ is reduced by transferring spectral weight {\em away} from the Fermi level. This is consistent with the fact, that $\sigma^{xx}$ is mainly determined by contributions from the Fermi level while $\sigma^{xy}$ is governed by the bands away from the Fermi level, as it has been discussed below Eq.~(\ref{equ:principalvaluecontribution}).
	
	\begin{figure}[t]
		\centering
		\includegraphics[width=0.9\linewidth]{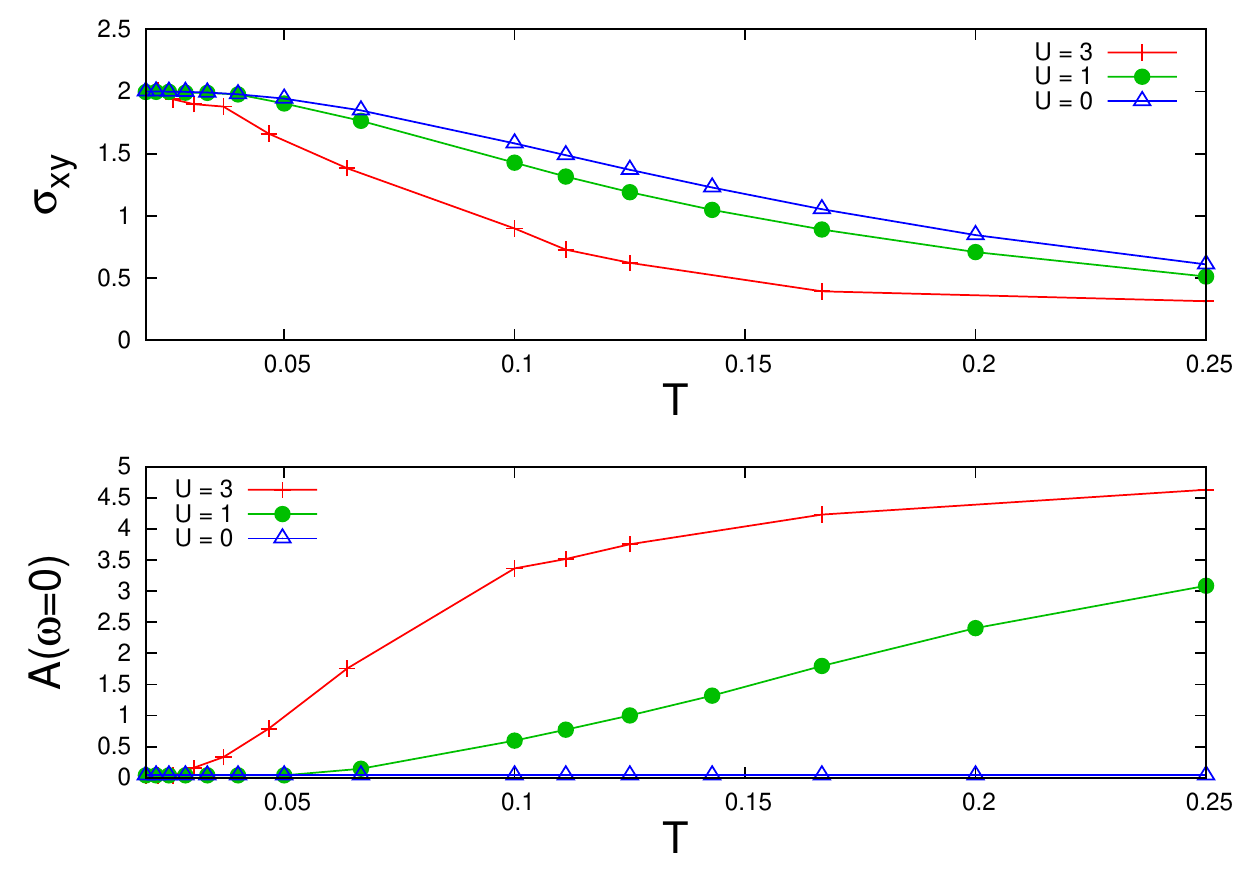}
		\caption{The temperature dependencies of the Hall Conductivity $\sigma_{xy}$ and the spectral function $A(\omega=0)$.}
		\label{fig:hallvsT}
	\end{figure}
	
	Let us finally address the temperature dependence of the Hall conductivity w.r.t. to $A(\omega=0)$ as depicted in Fig.~\ref{fig:hallvsT}. In the lower panel, one can clearly see, that the rate at which the first gap in the system is filled with temperature strongly depends on $U$. Again, this is well reflected in a corresponding suppression of $\sigma^{xy}$ and eventually a deviation from the integer values. This demonstrates that for intermediate-to-low temperatures a {\sl combined} effect of the temperature itself [given by the Fermi functions in Eq.~(\ref{equ:calcoptDMFT})] and the shift of spectral weight due to interactions [represented by the spectral functions $A_{l}(\omega,\mathbf{k})$ in Eq.~(\ref{equ:calcoptDMFT})] is responsible for the destruction of the integer quantum Hall state and the corresponding suppression of $\sigma^{xy}$.
	
	%As it was described in the section \ref{top} the main tendencies are change of the gap's widths and positions. Gaps become narrower with the interactions strength growth and at $U=2.5$ plateaus correspondent to the Integer Hall Effect seem to be close to disappear. Another feature of the interacting system appearing at large $U$ is the interval of incompressibility at half-filling correspondent to the Mott phase in our system.
 
	%$\bullet$ Fixed effective shift vs fixed $\langle n \rangle$
	
	%$\bullet$ Effects of non-zero temperatures and correlations: filling of the gap.

	%\subsection{$B = \frac{1}{3}$. Temperature dependencies}
	
	%\subsection{$B = \frac{1}{7}$}
	
	\section{\label{conclusions}Conclusion}
	
	We have studied the dc Hall conductivity $\sigma^{xy}$ in the $2d$ Hubbard model on a square lattice in a uniform transverse magnetic field at finite temperatures $T$ and Hubbard interactions $U$ in the framework of DMFT. In the noninteracting case at $T\!=\!0$, $\sigma^{xy}$ corresponds to an integer topological invariant of the magnetic band structure of the system, i.e., the first Chern number $\mathcal{C}_1$. We have analyzed, how the combined effect of finite temperature and correlations can modify this picture and were able to identify two main effects: (i) First, we have observed a shrinking of the size of the integer plateaus upon increasing $U$ w.r.t. the noninteracting case. This has been related one-to-one to the correlation driven renormalization of the band structure and, consequently, the reduction of the band gap width. (ii) Second, we have shown that for large values of the interaction the gap gets gradually filled  which leads to a reduction of $\sigma^{xy}$ from its integer values and, hence, to a breakdown of the integer quantum Hall regime.
	
	Moreover, we have observed an anomalous change of sign in $\sigma^{xy}$ for fillings just before the Mott transition (at half filling). This confirms previous studies \cite{rojo1993sign} where such a change of charge carriers from electrons to holes has been attributed to an correlation-driven anomaly of the kinetic energy as a function of the particle density close to the Fermi level.
	
	In this work, he have neglected the effect of the magnetic field on the spin degrees of freedom. It is indeed a very interesting question for future research work how the inclusion of Zeeman splitting will modify the physical picture outlined in this paper. Another natural question concerns the improvement of the method which has been used for the calculations. In fact, DMFT can provide only a purely frequency dependent self-energy which clearly cannot change the topological structure of the bands in momentum space. An extension of the present study by means of diagrammatic extension of DMFT \cite{Rohringer2018}, which include nonlocal correlation effects via a $\mathbf{k}$-dependent self-energy, will allow for a description of renormalization effects of the bands and represents, hence, a very promising future research direction.
	
	%How our description of the model is affected be including in the consideration of non-local correlations and interplay between Zeeman splitting and Mott physics can be a subject of future research.
	
	\section{Acknowledgments}
	
	We thank A. Toschi and G. Sangiovanni for interesting and useful discussions. The work is partly funded by Russian Science Foundadtion, grant  16-42-01057. A.M. is also supported by the ''Basis'' foundation under grant $\#$18-3-01.

	%$\bullet$ Half-filling and Mott transition.
	
	%\bullet$ Breakdown of the Integer regime due to local correlations and finite temperatures in case ($T<< \Delta \epsilon$).
	
	%$\bullet$ Zeeman splitting
	
	%$\bullet$ Non-local correlations
	
	\appendix
	
	\section{Vanishing of vertex correction for $\sigma_{xy}$ within DMFT}
	\label{app:novertcorr}
	
	In this Appendix, we present an explicit proof for the vanishing of vertex corrections in the calculation of $\sigma^{xy}(\nu)$ for the Hubbard model in a magnetic field within DMFT. Our considerations rely on a main feature of the model itself as well as on the properties of the DMFT approach which has been used to solve it: (i) the gauge invariance of the Hamiltonian and (ii) the locality of the self-energy $\Sigma$ and, in particular, of the irreducible (charge) vertex $\Gamma_{\text{ch}}$ within DMFT \cite{Zlatic1990}. We will pursue the following strategy for our proof: (i) We will perform the calculation of $\sigma^{xy}(\nu)$ for two different gauges and then (ii) show they yield the same result {\em only} for the bubble contribution while for terms including vertex corrections they differ in sign (and, hence, have to be zero).
	
	In a first step, let us introduce an alternative version of the Landau gauge, i.e., $\mathbf{\widetilde{A}}(\mathbf{r})=B(0,x,0)$. The corresponding dispersion relations and Fermi velocities are related as
	\begin{subequations}
	\label{equ:gauge}
	\begin{align}
	\label{equ:epsgauge}
	&\widetilde{\varepsilon}_{ll'}(k_x,k_y)=\varepsilon_{ll'}(-k_y,k_x),\\
	\label{equ:fermivelocityxgauge}
	&\mathbf{\widetilde{v}}^x_{ll'}(k_x,k_y)=\mathbf{v}^y_{ll'}(-k_y,k_x)\\
	\label{equ:fermivelocityygauge}
	&\mathbf{\widetilde{v}}^y_{ll'}(k_x,k_y)=-\mathbf{v}^x_{ll'}(-k_y,k_x).
	\end{align}
	\end{subequations}
	
	Next, we rewrite the term in the second line in Eq.~(\ref{equ:currentcurrentgeneral}) in order to dissect it into bubble contributions and terms containing vertex corrections:
	\begin{align}
	\label{equ:dissect2pgf}
	\sum_{\sigma'}\int_0^\beta e^{i\nu_n\tau}&\left\langle\mathcal{T}c^\dagger_{\mathbf{k}l_1\sigma}(\tau)c_{\mathbf{k}l_2\sigma}(\tau)c^\dagger_{\mathbf{k'}l_3\sigma'}(0)c_{\mathbf{k'}l_4\sigma'}(0)\right\rangle=\nonumber\\&=\frac{1}{\beta^2}\sum_{\omega_n\omega_n'}G^{(2),\omega_n\omega_n'\nu_n}_{l_1l_2l_3l_4,\text{ch}}(\mathbf{k},\mathbf{k'},\mathbf{q=0}),
	\end{align}
	where (taking into account SU(2) spin symmetry) $G^{(2)}_{\text{ch}}\!=\!G^{(2)}_{\uparrow\uparrow}\!+\!G^{(2)}_{\uparrow\downarrow}\!=\!G^{(2)}_{\downarrow\uparrow}\!+\!G^{(2)}_{\downarrow\downarrow}$ is the two-particle Green's function in the (ch)arge channel and $\omega_n^{(\prime)}\!=\!(2n^{(\prime)}+1)\frac{\pi}{\beta},n^{(\prime)}\in\mathds{Z}$, denotes fermionic Matsubara frequencies [the additional sum over $\sigma$ in Eq.~(\ref{equ:currentcurrentgeneral}) yields just an overall factor of $2$]. In the next step, we will split $G^{(2)}_{\text{ch}}$ into its different diagrammatic contributions, i.e., into bubble terms and a term containing the vertex corrections \cite{rohringer2012local}:
	\begin{align}
	\label{equ:splitG2}
	G^{(2),\omega_n\omega_n'\nu_n}_{l_1l_2l_3l_4,\text{ch}}(\mathbf{k},\mathbf{k'},&\mathbf{q=0})=2G_{l_2l_1}(i\omega_n,\mathbf{k})G_{l_4l_3}(i\omega_n',\mathbf{k'})\delta_{\nu_n 0}\nonumber\\-G_{l_4l_1}(i\omega_n,\mathbf{k})&G_{l_2l_3}(i\omega_n+i\nu_n,\mathbf{k})\delta_{\omega_n\omega_n'}\delta_{\mathbf{k}\mathbf{k'}}\nonumber\\-G_{\bar{l}_1l_1}(i\omega_n,\mathbf{k})&G_{l_2\bar{l}_2}(i\omega_n+i\nu_n,\mathbf{k})F_{\bar{l}_1\bar{l}_2\bar{l}_3\bar{l}_4,\text{ch}}^{\omega_n\omega_n'\nu_n}(\mathbf{k},\mathbf{k'},\mathbf{q=0})\nonumber\\&\times G_{\bar{l}_3l_3}(i\omega_n',\mathbf{k'})G_{l_4\bar{l}_4}(i\omega_n'+i\nu_n,\mathbf{k'}),
	\end{align}
	where the Einstein summation convention has been adopted for the orbital indices $\bar{l}_i$. $F_{\text{ch}}$ denotes the full (charge) vertex of the system which can be expressed by using the Bethe-Salpeter equations \cite{rohringer2012local} in in terms of the irreducible vertex $\Gamma_{\text{ch}}$:
	\begin{widetext}
	\begin{align}
	\label{equ:BS}
	F_{\bar{l}_1\bar{l}_2\bar{l}_3\bar{l}_4,\text{ch}}^{\omega_n\omega_n'\nu_n}&(\mathbf{k},\mathbf{k'},\mathbf{0})=\Gamma_{\bar{l}_1\bar{l}_1\bar{l}_1\bar{l}_1,\text{ch}}^{\omega_n\omega_n'\nu_n}\delta_{\bar{l}_1\bar{l}_2\bar{l}_3\bar{l}_4}+\frac{\delta_{\bar{l}_1\bar{l}_2}}{\beta}&\sum_{\bar{\omega}_n\mathbf{\bar{k}}}\sum_{\bar{l}_5\bar{l}_6}\Gamma_{\bar{l}_1\bar{l}_1\bar{l}_1\bar{l}_1,\text{ch}}^{\omega_n\bar{\omega}_n\nu_n}G_{\bar{l}_5\bar{l}_1}(\bar{i\omega}_n,\mathbf{\bar{k}})G_{\bar{l}_1\bar{l}_6}(i\bar{\omega}_n+i\nu_n,\mathbf{\bar{k}}) F_{\bar{l}_5\bar{l}_6\bar{l}_3\bar{l}_4,\text{ch}}^{\bar{\omega}_n\omega_n'\nu_n}(\bar{\mathbf{k}},\mathbf{k'},\mathbf{0}).
	\end{align}
	\end{widetext}
	where we have taken into account that $\Gamma_{\text{ch}}$ is purely local within DMFT, i.e., it does not depend on the momenta and is also diagonal in {\em all} orbital indices, since they just represent different lattice sides within the magnetic unit cell. The right hand side of Eq.~(\ref{equ:splitG2}) does not depend on $\mathbf{k}$ and, hence, $F_{\text{ch}}$ is independent of this momentum. Moreover, the right hand side is proportional to $\delta_{\bar{l}_1\bar{l}_2}$ and so has to be the left hand side. Finally, in the sum on the right hand side of Eq.~(\ref{equ:splitG2}), $F_{\text{ch}}$ and $\Gamma_{\text{ch}}$ can be exchanged which means that $F_{\text{ch}}$ is also independent of $\mathbf{k'}$ and proportional to $\delta_{\bar{l}_3\bar{l}_4}$. This leads to the following simplification of the term in the third line of Eq.~(\ref{equ:splitG2}) (i.e., the vertex corrections):
	\begin{widetext}
	\begin{align}
	\label{equ:vertexcorrsimple}
	&-G_{\bar{l}_1l_1}(i\omega_n,\mathbf{k})G_{l_2\bar{l}_2}(i\omega_n+i\nu_n,\mathbf{k})F_{\bar{l}_1\bar{l}_2\bar{l}_3\bar{l}_4,\text{ch}}^{\omega_n\omega_n'\nu_n}(\mathbf{k},\mathbf{k'},\mathbf{q=0}) G_{\bar{l}_3l_3}(i\omega_n',\mathbf{k'}) G_{l_4\bar{l}_4}(i\omega_n'+i\nu_n,\mathbf{k'})\nonumber\\=&-G_{ll_1}(i\omega_n,\mathbf{k})G_{l_2l}(i\omega_n+i\nu_n,\mathbf{k})F_{ll',\text{ch}}^{\omega_n\omega_n'\nu_n}G_{l'l_3}(i\omega_n',\mathbf{k'})G_{l_4l'}(i\omega_n'+i\nu_n,\mathbf{k'}),
	\end{align}
	\end{widetext}
	where we have adopted the notation $\bar{l}_1\!=\!\bar{l}_2\!=\!l$ and $\bar{l}_3\!=\!\bar{l}_4\!=\!l'$. Let us point out that $F_{ll',\text{ch}}^{\omega_n\omega_n'\nu_n}$ fulfills the symmetry relation
	\begin{align}
	\label{equ:symmetry}
	F_{ll',\text{ch}}^{\omega_n\omega_n'\nu_n}\!=\!F_{l'l,\text{ch}}^{\omega_n'\omega_n\nu_n}.
	\end{align}
	This can be seen in the following way: First, the irreducible vertex of the AIM {\sl does not} include a magnetic field and is, hence, time reversal invariant which is expressed by the relation \cite{rohringer2012local} $\Gamma_{\text{ch}}^{\omega_n\omega_n'\nu_n}\!=\!\Gamma_{\text{ch}}^{\omega_n'\omega_n\nu_n}$. The symmetry w.r.t. the $l_i$'s can be seen most easily by iterating Eq.~(\ref{equ:BS}). The Green's functions inside the sum then become $G_{\bar{l}_4\bar{l}_1}(\bar{i\omega}_n,\mathbf{\bar{k}})G_{\bar{l}_1\bar{l}_4}(i\bar{\omega}_n+i\nu_n,\mathbf{\bar{k}})$. If we exchange $\bar{l}_1\!\leftrightarrow\!\bar{l}_3$ and $\bar{l}_2\!\leftrightarrow\!\bar{l}_4$ (which corresponds to the exchange $l\!\leftrightarrow\!l'$ in Eq.~(\ref{equ:symmetry}) we obtain (considering that $\bar{l}_1\!=\!\bar{l}_2$ and $\bar{l}_3\!=\!\bar{l}_4$) $G_{\bar{l}_1\bar{l}_4}(\bar{i\omega}_n,\mathbf{\bar{k}})G_{\bar{l}_4\bar{l}_1}(i\bar{\omega}_n+i\nu_n,\mathbf{\bar{k}})$. Due to the special structure of the matrix $G_{ll'}$ in orbital space we have the relation $G_{ll'}(i\omega_n,\bar{k}_x,\bar{k}_y)\!=\!G_{l'l}(i\omega_n,\bar{k}_x,-\bar{k}_y)$, i.e., transposition of this matrix changes only the sign of $\bar{k}_y$. This sign change, however, can be easily compensated by a transformation of the integration variable $\bar{k}_y\!\rightarrow\!-\bar{k}_y$ which proves Eq.~(\ref{equ:symmetry}) up to the $2^{\text{nd}}$ order in $\Gamma_{\text{ch}}$. The extension to higher orders is obvious.

	We will now proceed with the calculation of $\chi^{xy}(i\nu_n)$ where we consider the three different contributions to $G^{(2)}_{\text{ch}}$ [in the first, second and third line of Eq.~(\ref{equ:splitG2}), respectively] separately. To this end, let us first note that the contribution in the first line of Eq.~(\ref{equ:splitG2}) is proportional to $\delta_{\nu_n0}$. Hence, if we perform the analytic continuation $\nu_n\!\rightarrow\!0$ this term vanishes.
	
	As for the contribution in the second line of Eq.~(\ref{equ:splitG2}), we obtain
	\begin{align}
	\label{equ:chibubble}
	\chi^{xy}_{(2)}(i\nu_n)=\frac{2e^2}{\hbar V}&\sum_{\omega_n\mathbf{k}}\sum_{l_1l_2l_3l_4}v_{\mathbf{k},l_1l_2}^{x}v_{\mathbf{k},l_3l_4}^{y}\nonumber\\&\times G_{l_4l_1}(i\omega_n,\mathbf{k})G_{l_2l_3}(i\omega_n+i\nu_n,\mathbf{k}).
	\end{align}
	Let us now perform the same calculation using the alternative gauge $\mathbf{\widetilde{A}}(\mathbf{r})$ which yields:
	\begin{align}
	\label{equ:chibubblegauge}
	\widetilde{\chi}^{xy}_{(2)}(i\nu_n)=\frac{2e^2}{\hbar V}&\sum_{\omega_n\mathbf{k}}\sum_{l_1l_2l_3l_4}\widetilde{v}_{\mathbf{k},l_1l_2}^{x}\widetilde{v}_{\mathbf{k},l_3l_4}^{y}\nonumber\\&\times \widetilde{G}_{l_4l_1}(i\omega_n,\mathbf{k})\widetilde{G}_{l_2l_3}(i\omega_n+i\nu_n,\mathbf{k}),
	\end{align}
	where $\widetilde{G}$ is the Green's function constructed from $\widetilde{\varepsilon}_{ll'}(\mathbf{k})$ in Eq.~(\ref{equ:epsgauge}). We can now recast Eq.~(\ref{equ:chibubblegauge}) into the form of Eq.~(\ref{equ:chibubble}) by means of the following steps: (i) First, using Eqs.~(\ref{equ:gauge}), we can replace all quantities in the new gauge ($\widetilde{x}$) with the corresponding quantities in the original gauge ($x$). This introduces an additional minus sign for the $k_y$ variable which, however, can be readily eliminated by the variable transformation $k_y\!\rightarrow\!-k_y$ inside the $\mathbf{k}$-sum. Next, we can also exchange $k_x$ with $k_y$ which leads to
	\begin{align}
	\label{equ:chibubblegaugerewritten}
	\widetilde{\chi}^{xy}_{(2)}(i\nu_n)=-\frac{2e^2}{\hbar V}&\sum_{\omega_n\mathbf{k}}\sum_{l_1l_2l_3l_4}v_{\mathbf{k},l_1l_2}^{y}v_{\mathbf{k},l_3l_4}^{x}\nonumber\\&\times G_{l_4l_1}(i\omega_n,\mathbf{k})G_{l_2l_3}(i\omega_n+i\nu_n,\mathbf{k}),
	\end{align}
	i.e., we obtain an additional minus sign w.r.t. Eq.~(\ref{equ:chibubble}) and the Fermi velocities $x$- and $y$-directions are interchanged. We can now perform the index transformation $l_1\!\leftrightarrow\!l_3$ and $l_2\!\leftrightarrow\!l_4$ to obtain
	\begin{align}
	\label{equ:chibubblegaugerewritten2}
	\widetilde{\chi}^{xy}_{(2)}(i\nu_n)=-\frac{2e^2}{\hbar V}&\sum_{\omega_n\mathbf{k}}\sum_{l_1l_2l_3l_4}v_{\mathbf{k},l_1l_2}^{x}v_{\mathbf{k},l_3l_4}^{y}\nonumber\\&\times G_{l_4l_1}(i\omega_n+i\nu_n,\mathbf{k})G_{l_2l_3}(i\omega_n,\mathbf{k}).
	\end{align}
	Finally, we transform the fermionic frequency as $\omega_n\!\rightarrow\!\omega_n\!-\!\nu_n$ which yields that
	\begin{align}
	\label{equ:relatechibubblechibubblegauge}
	\widetilde{\chi}^{xy}_{(2)}(i\nu_n)=-\chi^{xy}_{(2)}(-i\nu_n).
	\end{align}
	For the limit $i\nu_n\!\rightarrow\nu\!+\!i\delta$ [see Eqs.~(\ref{equ:calcoptDMFT}) and (\ref{equ:principalvaluecontribution}) in the main text], the relevant part of $\chi^{xy}$ is proportional to $\nu_n$ and, hence, the minus sign in front of $\chi^{xy}$ is canceled. One can see this also more directly by going through the steps in Eqs.~(\ref{equ:calcoptDMFT})-(\ref{equ:principalvaluecontribution}) for both gauges. In the final expression (\ref{equ:principalvaluecontribution}), the $v^x$ and $v^y$ are then exchanged for the $\mathbf{\widetilde{A}}$-gauge and one has an additional minus sign due to Eq.~(\ref{equ:fermivelocityygauge}). In order to restore the original order, one can use the cyclic property of the trace which however leads then to an exchange of the order of $\omega$ and $\omega'$. This can be compensated by an additional exchange of the Fermi functions in the first line of Eq.~(\ref{equ:principalvaluecontribution}) which yields a further minus sign which cancels the first one. This shows that (after the before mentioned manipulations) we arrive at the exactly same expression for $\chi^{xy}_{(2)}$ in both gauges.
	
	\subsection{The vertex correction contribution $\chi^{xy}_v(i\nu_n)$}
	\label{sec:vertexcorrection}
	
	Let us now calculate the contribution to of the vertex part [see Eq.~(\ref{equ:vertexcorrsimple})] to $\chi^{xy}$:
	\begin{align}
	\label{equ:chivertexpart}
	&\chi^{xy}_v(i\nu_n)=\frac{2e^2}{\hbar V}\sum_{\omega_n\omega_n'}\sum_{\mathbf{k}\mathbf{k'}}\sum_{l_ill'}v_{\mathbf{k},l_1l_2}^{x}v_{\mathbf{k},l_3l_4}^{y}G_{ll_1}(i\omega_n,\mathbf{k})\nonumber\\&\times G_{l_2l}(i\omega_n+i\nu_n,\mathbf{k})F_{ll',\text{ch}}^{\omega_n\omega_n'\nu_n}G_{l'l_3}(i\omega_n',\mathbf{k'})G_{l_4l'}(i\omega_n'+i\nu_n,\mathbf{k'}).
	\end{align}
	The corresponding expression in the other gauge $\mathbf{\widetilde{A}}$ reads:
	\begin{align}
	\label{equ:chivertexpartgauge}
	&\widetilde{\chi}^{xy}_v(i\nu_n)=\frac{2e^2}{\hbar V}\sum_{\omega_n\omega_n'}\sum_{\mathbf{k}\mathbf{k'}}\sum_{l_ill'}\widetilde{v}_{\mathbf{k},l_1l_2}^{x}\widetilde{v}_{\mathbf{k},l_3l_4}^{y}\widetilde{G}_{ll_1}(i\omega_n,\mathbf{k})\nonumber\\&\times\widetilde{G}_{l_2l}(i\omega_n+i\nu_n,\mathbf{k}) F_{ll',\text{ch}}^{\omega_n\omega_n'\nu_n}\widetilde{G}_{l'l_3}(i\omega_n',\mathbf{k'})\widetilde{G}_{l_4l'}(i\omega_n'+i\nu_n,\mathbf{k'}).
	\end{align}
	We can express $\widetilde{\chi}^{xy}$ in terms of quantities in the original gauge by using Eqs.~(\ref{equ:gauge}) and performing the variable transformations $k^{(\prime)}_y\!\rightarrow\!-k^{(\prime)}_y$ and $k_x^{(\prime)}\!\leftrightarrow\!k_y^{(\prime)}$ inside the $\mathbf{k}$-sums (note that $F_{ll',\text{ch}}$ is gauge invariant). This yields
	\begin{align}
	\label{equ:chivertexpartgauge1}
	&\widetilde{\chi}^{xy}_v(i\nu_n)=-\frac{2e^2}{\hbar V}\sum_{\omega_n\omega_n'}\sum_{\mathbf{k}\mathbf{k'}}\sum_{l_ill'}v_{\mathbf{k},l_1l_2}^{y}v_{\mathbf{k},l_3l_4}^{x}G_{ll_1}(i\omega_n,\mathbf{k})\nonumber\\&\times G_{l_2l}(i\omega_n+i\nu_n,\mathbf{k}) F_{ll',\text{ch}}^{\omega_n\omega_n'\nu_n}G_{l'l_3}(i\omega_n',\mathbf{k'}) G_{l_4l'}(i\omega_n'+i\nu_n,\mathbf{k'}).
	\end{align}
	We can now proceed with relabeling $l_1\!\leftrightarrow\!l_3$, $l_2\!\leftrightarrow\!l_4$, $l\!\leftrightarrow\!l'$, $\omega_n\!\leftrightarrow\!\omega_n'$ and $\mathbf{k}\!\leftrightarrow\!\mathbf{k'}$ which gives
	\begin{align}
	\label{equ:chivertexpartgauge2}
	&\widetilde{\chi}^{xy}_v(i\nu_n)=-\frac{2e^2}{\hbar V}\sum_{\omega_n\omega_n'}\sum_{\mathbf{k}\mathbf{k'}}\sum_{l_ill'}v_{\mathbf{k},l_1l_2}^{x}v_{\mathbf{k},l_3l_4}^{y}G_{ll_1}(i\omega_n,\mathbf{k})\nonumber\\&\times G_{l_2l}(i\omega_n+i\nu_n,\mathbf{k}) F_{l'l,\text{ch}}^{\omega_n'\omega_n\nu_n}G_{l'l_3}(i\omega_n',\mathbf{k'})G_{l_4l'}(i\omega_n'+i\nu_n,\mathbf{k'}).
	\end{align}
	The symmetry $F_{l'l,\text{ch}}^{\omega_n'\omega_n\nu_n}\!=\!F_{ll',\text{ch}}^{\omega_n\omega_n'\nu_n}$ [see Eq.~(\ref{equ:symmetry})] implies that
	\begin{align}
	\label{equ:chivertantisymm}
	\widetilde{\chi}^{xy}_v(i\nu_n)=-\chi^{xy}_v(i\nu_n),
	\end{align}
	and, hence, this term vanishes.
	
	\bibliography{QHE}

\end{document}